\documentclass[prx,superscriptaddress,twocolumn,floatfix,notitlepage,citeautoscript]{revtex4-2}
\usepackage[T2A]{fontenc}
\usepackage[cp1251]{inputenc}
\usepackage{amssymb}
\usepackage{amsmath}
\usepackage{mathrsfs}
\usepackage{graphicx}
\usepackage[labelfont=bf,format=plain,justification=centerlast,font=footnotesize]{caption}
\usepackage{array}
\usepackage{epstopdf}

\newcommand{\tr}{\mathop{\rm tr}\nolimits}
\newcommand{\diverg}{\mathop{\rm div}\nolimits}

\begin{document}

\title{Tunable Bose-Einstein condensation and roton-like excitation spectra\\ with dipolar exciton-polaritons in crossed fields}

\author{Timofey V.~Maximov}
\affiliation{National Research University ``Moscow Institute of Physics and Technology'', Phystech School of Fundamental and Applied Physics, 141700 Dolgoprudny, Moscow Region, Russia}
\affiliation{N.~L. Dukhov Research Institute of Automatics (VNIIA), Moscow, 127055, Russia}

\author{Igor~L. Kurbakov}%
\affiliation{Institute for Spectroscopy RAS, 142190 Troitsk, Moscow, Russia}%

\author{Nina~S.~Voronova}
\email{nsvoronova@mephi.ru}
\affiliation{National Research Nuclear University MEPhI (Moscow Engineering Physics Institute), 115409 Moscow, Russia}
\affiliation{Russian Quantum Center, Skolkovo IC, Bolshoy boulevard 30 bld. 1, 121205 Moscow, Russia}

\author{Yurii~E. Lozovik}%
\email{lozovik@isan.troitsk.ru}%
\affiliation{Institute for Spectroscopy RAS, 142190 Troitsk, Moscow, Russia}%
\affiliation{National Research University Higher School of Economics, 109028 Moscow, Russia}%

\begin{abstract}
We develop the many-body theory of dipolar exciton-polaritons in an optical microcavity in crossed transverse electric and in-plane magnetic fields. Even for relatively weak fields, we reveal the existence of two minima in the bare lower-polariton dispersion, which give rise to the tuneable transition between the polariton Bose-Einstein condensate and that of excitons, produced by the competition between these minima. We predict that such dipolar condensate exhibits a roton-maxon character of the excitation spectrum, never before observed for polaritons. We show that upon the transition between the two condensation regimes, the weak correlations in the polariton gas give way to the intermediate interparticle correlations characteristic for excitons, and that the transition is accompanied by a sharp quenching of photoluminescence as the lifetime is increased by several orders of magnitude. While in the polariton regime, the luminescence peak corresponding to the condensate is shifted to a non-zero angle. The angular dependence of the two-photon decay time in the Hanbury Brown and Twiss experiment is calculated and used as a tool to evidence the formation of the macroscopically-coherent state. Our proposal opens opportunities towards manipulating the superfluid properties and extended-range dipole-dipole correlations of exciton-polariton condensates.
\end{abstract}

\maketitle

\section{Introduction}

Dipole-dipole interactions are key to a variety
of many-body phenomena and various phases, both in Fermi and Bose gases, at temperatures low enough to achieve quantum degeneracy. Dipolar ultracold atomic systems have been shown to exhibit superfluid $p$-wave Cooper pairing~\cite{you,shlyap1}, rotons~\cite{kuzirski,shlyap2}, Mott-insulator and checkerboard phases~\cite{lewenstein,bloch}, and supersolid formation~\cite{giovanazzi,pfau,pfau_rev}. While such gases are weakly-interacting compared to liquid helium, their interactions are controllable by means of Feshbach resonance~\cite{ketterle} or external off-resonance laser fields~\cite{kuzirski}. In either of cases, the origin of formation of such exotic states lies in the momentum dependence of the interparticle interaction, which results in the roton-maxon spectrum of excitations~\cite{Landau,Feynman,shlyap3}. To this end, it is a general physical phenomenon that should be present in any interacting gas with an extended-range momentum dependence of the scattering amplitude.

In this context, another type of systems where interactions can be manipulated using external fields has been considered: that of excitons~\cite{snoke} and exciton-polaritons~\cite{microcavities}. The exciton---a neutral bound state of an electron and a hole in a semiconductor---due to its fermionic components can be made dipolar by applying electric field, which makes exciton gases easily tuneable, both with respect to interactions and their lifetime~\cite{lozovikIX,jetpl0840222}. The appearance of the dipole moment brings to cold exciton gases a plethora of many-body phenomena, including roton instabilities~\cite{prb090165430,shelykh1}, supersolidity~\cite{prl108060401}, density waves~\cite{prb091245302} and other phases~\cite{prl098060405}.
The exciton-polariton---a hybrid quasiparticle resulting from quantum-well excitons coupling to photons in an optical microcavity---possess additional degrees of freedom, such as the Rabi splitting, the photon-exciton energy detuning, and pseudospin. Compared to helium~\cite{pr0104000576}, atoms~\cite{sci269000198}, and excitons~\cite{jetpl0840329,butov2,dubin}, the polariton Bose-Einstein condensation (BEC) occurs at much higher temperatures~\cite{rmp085000299} due to extremely light effective mass inherited from cavity photons. The exciton component, on the other hand, provides polaritons with interactions. Yet, bringing the dipolar exciton physics to quantum-degenerate polariton systems has failed so far: due to the increased electron-hole separation, indirect excitons suffer from the quench of the oscillator strength, hence their Rabi coupling to light is reduced. As such, while dipolar polaritons (dipolaritons) have been observed both in GaAs coupled quantum wells (QWs)~\cite{baumberg} and, more recently, in MoS$_2$ homobilayers~\cite{menon,tartakovskii} by means of hybridizing them with the direct exciton, the realisation of dipolar polariton BEC remains elusive. The rotonization of polariton excitation spectrum has nevertheless been theoretically discussed, in the context of coupling the system to a two-dimensional (2D) electron gas~\cite{cotlet,sokolik} and accounting for their spin degree of freedom~\cite{shelykh2}.

Here, we consider a different setting to study dipolariton Bose condensation and excitation spectrum. In particular, instead of focusing on a double-layer systems like coupled QWs or transition-metal dichalcogenide bilayers, we demonstrate that the strong-coupling regime can be preserved in a wide single QW embedded in a microcavity, in the presence of relatively weak transverse electric fields (so as to create the exciton dipole, at the same time not precluding the polariton BEC formation). Notably, the BEC of dipolar excitons (without coupling to light) in such wide QWs has previously been realized~\cite{jetpl0840329}.
Furthermore, we show that a fine control over the single-particle dispersion, interparticle interactions, and excitation spectrum can be acquired when one additionally applies magnetic field directed in the QW plane (a schematic illustration is provided in Fig.~\ref{fig1}{\bf a}).

The influence of magnetic field on electrically-charged constituents of an exciton has been studied both experimentally~\cite{prb062001548,prb062015323,prl087216804,jetpl0890019,jetpl0890510} and theoretically~\cite{gorkov,ruvinskii_pla,tokatly_ssc,prb065235304}. The Lorentz force acting on the electron and hole breaks both the time-reversal and space-inversion symmetries; as a result, the exciton momentum becomes an irrelevant quantity, giving way to {\it magnetic momentum} as the new integral of motion. In transverse magnetic field, there is a possibility of magnetoexciton formation due to the competition of the exciton hydrogen-like energy states with the magnetic-field Landau levels~\cite{gorkov,ruvinskii_pla,prb065235304}. When the magnetic field has an in-plane component, the paraboloid exciton dispersion $p^2/2m_{\rm ex}$, whose intersection with the light cone is dictated solely by the exciton effective mass $m_{\rm ex}$, shifts to $\propto ({\bf p-p}_0)^2$, where the displacement momentum ${\bf p}_0$ lies in the plane of the QW perpendicularly to the magnetic field and is defined via the product of the field strength and the exciton dipole moment~\cite{tokatly_ssc,prb062001548,prb062015323}.

The aim of this work is to study the quantum-coherent properties of wide-QW dipolaritons in crossed (transverse electric and in-plane magnetic) fields, where the described effect of the fields on the exciton dispersion is combined with strong coupling to the electromagnetic mode inside the cavity.
Our theory predicts the existence of two minima on the lower branch of the polariton dispersion, which can be tuned at fixed fields strengths by a purely polaritonic control parameter, namely the photon-exciton detuning. When macroscopic occupation of the ground state (now differing from ${\bf p}=0$) is considered, we show that the competition of these two minima in energy brings about remarkable effects completely new for polariton physics. In particular, the transition between the polariton and exciton BEC regimes (and vice versa) is achieved by tuning the system parameters and is accompanied by suppression of losses by several orders of magnitude. We show that this transition between the two BECs displays features of a first-order phase transition. For both regimes, the spectrum of elementary excitations is asymmetric and features pronounced, controllable roton minima.
We address the exciton features, such as interparticle correlations~\cite{prb087205302,ssc144000399,prb080195313} and extended range of the dipole-dipole pair potential~\cite{prb080195313,prb095245430}, and, on the other hand, the polariton specifics, in particular the absense of Galilean invariance~\cite{semenov}, non-parabolicity of the dispersion, and the presence of the momentum-dependent Hopfield weights 
in the two-body and many-body interactions. We discuss the implications of the absence of central symmetry and parity with respect to momentum, and provide the conditions of the system stability.
Finally, we calculate the two-photon coherence in the Hanbury Brown and Twiss (HBT) setting~\cite{jetpl0900146,prb081235402} and outline the means to evidence the dipolariton BEC formation.

The paper is organised as follows. In Section~\ref{ham}, we introduce the system, solve the dipolar exciton eigenvalue problem in crossed fields, and discuss the influence of the electric field on the exciton dipole moment and the strength of their coupling to light. We derive the effective Hamiltonian dressed with extended-range exciton-exciton interactions, obtain the bare polariton dispersion, and analyze its shape dependent on the control parameters of the system. In Sec.~\ref{B-approx}, we develop the Bogoliubov apparatus accounting for the fact that the ground state (and hence the macroscopically occupied state) of the system corresponds to a non-zero in-plane momentum and study the nature of the transition between the two BEC regimes. Sec.~\ref{sec_corr} is devoted to calculation of various correlators, such as the polariton occupation number and their one-body density matrix, the momentum-frequency distribution of excitations, and the condensate fraction in the system. The anomalous Green's function, luminescence intensity distributions, and two-photon HBT coincidence experiment are discussed in Sec.~\ref{B-opt}. Sec.~\ref{Concl} summarizes our findings. The details of some derivations are provided in Appendices~\ref{AppH}, \ref{AppMin}, and \ref{AppU}. Appendix~\ref{AppDet} is devoted to the description of the transition when changing the detuning instead of density.

\section{Wide-quantum-well dipolaritons}\label{ham}

\begin{figure}[t]
\includegraphics[width=0.9\linewidth]{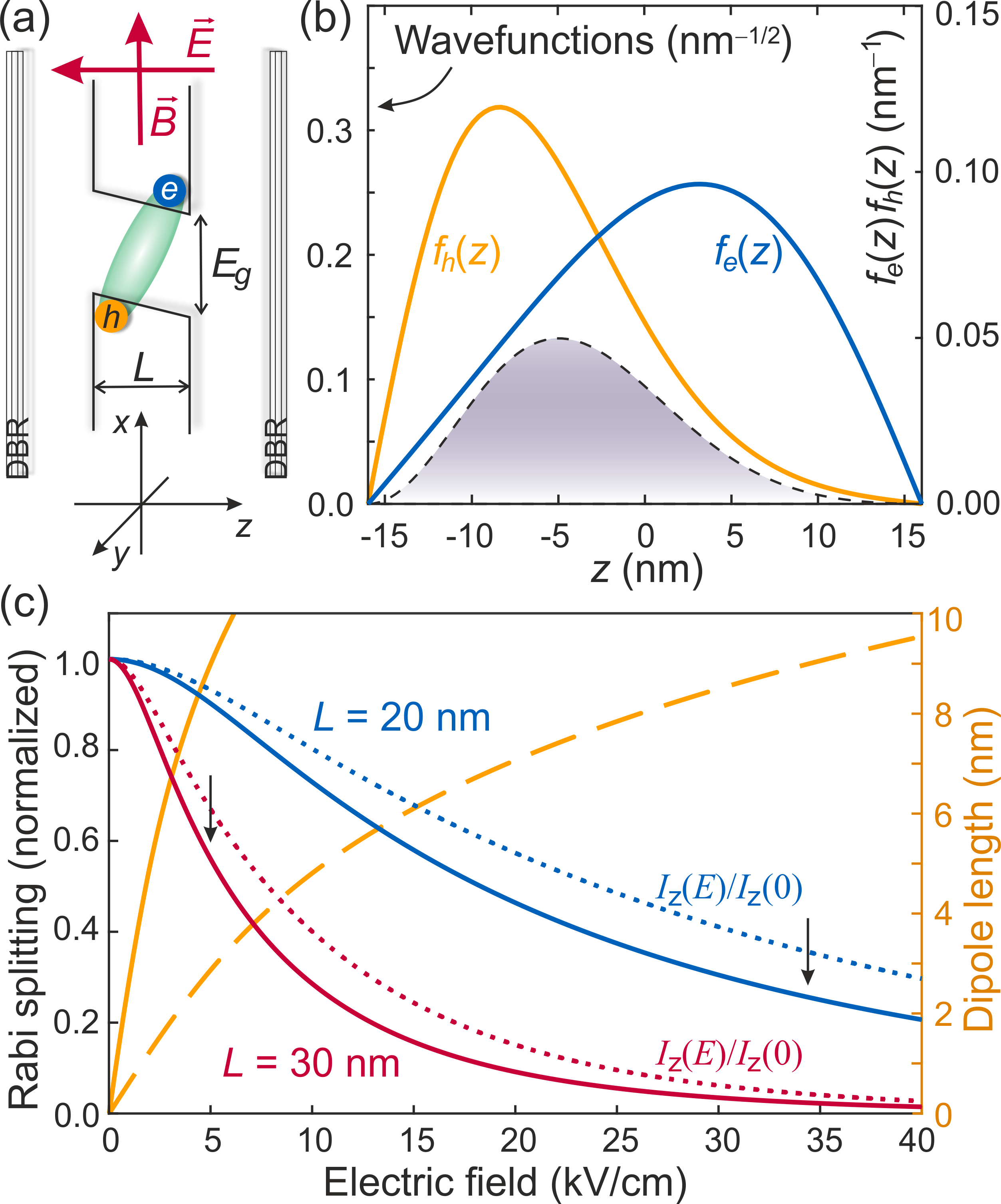}
\caption{(а) Schematic representation of a wide QW of the width $L$ in cross electric and magnetic fields inside a microcavity (not in scale). (b) Electron (blue) and hole (yellow) wavefunctions (the left axis) in a 30~nm GaAs QW according to the solution of Eq.~(\ref{Schred}) at $E=5$~kV/cm, $B=0$, $L=30$~nm, and their product (the right axis, dashed black line) dependent on the transverse coordinate $z$. (c) Left axis: the
transverse $I_z(E)$ overlap integral (dotted lines) and
the Rabi splitting $\hbar\Omega_0$ (solid lines) dependent on the electric field $E$ at $B=0$, normalized to their values at $E=0$.
The blue and red lines correspond to the QW width $L=20$~nm and 30~nm, respectively.
Right axis: the exciton dipole length $d/e$ versus electric field $E$. The solid line for $L=30$~nm, the dashed line for $L=20$~nm.
The black arrows denote the fields and the Rabi splitting values corresponding to the dipole length of 9~nm.
}
\label{fig1}
\end{figure}

The starting point of our discussion is the Hamiltonian of wide-QW excitons interacting with light in presence of static in-plane magnetic field $B{\bf e}_x$ and transverse electric field $-E\vec{\mathrm{e}}_z$ (see the sketch in Fig.~\ref{fig1}{\bf a}):
\begin{multline}\label{Hexph}
\hat{\mathscr{H}} \!=\!\! \sum\limits_{\bf p}\!E_{\bf p}\hat{Q}_{\bf p}^\dag\hat{Q}_{\bf p} + \!\sum\limits_{\bf p}\!
\hbar\omega_{\bf p}\hat{c}_{\bf p}^\dag\hat{c}_{\bf p} + \frac{1}{2}\!\sum\limits_{\bf p}\!
\left[\hbar\Omega_{\bf p}\hat{Q}_{\bf p}^\dag\hat{c}_{\bf p} \!+ \text{h.c.}\right] \\
+ \frac{1}{2}\sum\limits_{\bf p,q,q^\prime} U_0({\bf p,q,q^\prime})
\hat{Q}_{\bf q}^\dag\hat{Q}_{\bf q^\prime}^\dag\hat{Q}_{\bf q^\prime+p}\hat{Q}_{\bf q-p}.
\end{multline}
Throughout the paper, we will use the arrowhead notation for three-dimensional (3D) vectors having an out-of-plane component, while boldface is chosen to denote 2D vectors on the $(x,y)$--plane. In Eq.~(\ref{Hexph}), $\hat{Q}_{\bf p}$ is the annihilation operator of an exciton with the in-plane momentum ${\bf p}$, the wavefunction $\phi_{\bf p}(\vec{r}_e,\vec{r}_h)$ and dispersion $E_{\bf p}$ which we define below [$\vec{r}_{e(h)}\equiv\{{\bf r}_{e(h)},z_{e(h)}\}$ are the electron (hole) 3D position vectors within the QW, see Fig.~\ref{fig1}{\bf a}]. The annihilation operator of a cavity photon with the momentum ${\bf p}$ is denoted as $\hat{c}_{\bf p}$, with the single-particle dispersion $\hbar\omega_{\bf p}=(E_{\rm ph}^2 + p^2c^2/\varepsilon)^{1/2}\approx E_{\rm ph}+ p^2/2m_{\rm ph}$, where $E_{\rm ph}$ is the cavity ground state and $m_{\rm ph}=E_{\rm ph}\varepsilon/c^2$ denotes the photon effective mass, $c$ is the velocity of light in vacuum, $\varepsilon$ the dielectric constant of the medium.
Summation over the polarization (spin projection) index is omitted, as we consider here only the mode in which the resulting polaritons experience Bose condensation.

The third term in (\ref{Hexph}) describes the light-matter coupling with
the Rabi splitting
\begin{multline}\label{OmegaRabi}
\hbar\Omega_{\bf p} =\!
\left|E_g\sqrt{\frac{8\pi}{\varepsilon\hbar\omega_{\bf p}S}}
(\vec{\rm e}_{\bf p}\vec{\rm d}_{\rm vc})
\!\! \int \!\!d\vec{r}_ed\vec{r}_h e^{i{\bf p\cdot r}/\hbar}\varphi(z) \times \right. \\
\Biggl. \phi^*_{\bf p}(\vec{r}_e,\vec{r}_h)
\delta(\vec{r}_e-\vec{r}_h)\Biggr|,
\end{multline}
where $E_g$ is the semiconductor gap energy, $S$ is the area of quantization, $\vec{\rm e}_{\bf p}$ is the polarization vector, $\vec{\rm d}_{\rm vc}$ is the interband dipole, and $\varphi(z)$ is the transverse-quantized photon wavefunction.
The last term in (\ref{Hexph}) describes the bare direct pair interaction of excitons, with the Fourier image of the potential
\begin{align}
U_0({\bf p}, {\bf q}, {\bf q}^\prime) =
\!\int\!\! d\vec{r}_e d\vec{r}_h d\vec{s}_e d\vec{s}_h U_0(\vec{r}_e,\vec{r}_h,\vec{s}_e,\vec{s}_h) \times \nonumber\\
\phi_{\bf q}^*(\vec{r}_e,\vec{r}_h)\phi_{\bf q^\prime}^*(\vec{s}_e,\vec{s}_h)
\phi_{{\bf q}^\prime + {\bf p}}(\vec{r}_e,\vec{r}_h) \phi_{{\bf q} - {\bf p}}(\vec{s}_e,\vec{s}_h), \label{U0pkk'}
\end{align}
where $U_0(\vec{r}_e,\vec{r}_h,\vec{s}_e,\vec{s}_h)$ is the potential of the direct Coulomb interaction of an electron and a hole belonging to different exciton species.
The derivation of this Hamiltonian from the electron-hole picture, accounting for interaction of the system with electromagnetic field inside a microcavity, is provided in Appendix~\ref{AppH}.

\subsection{The exciton eigenvalue problem}

The main interest of the Hamiltonian (\ref{Hexph}) before its diagonalization is represented by the exciton single-particle dispersion $E_{\bf p}$ and wavefunction $\phi_{\bf p}(\vec{r}_e,\vec{r}_h)$ in crossed fields. They are defined from the eigenvalue problem
$\hat H_0(\vec{r}_e,\vec{r}_h)\phi_{\bf p}(\vec{r}_e,\vec{r}_h) = E_{\bf p}\phi_{\bf p}(\vec{r}_e,\vec{r}_h)$,
with
\begin{multline}\label{Hrerh}
\hat{H}_0 \!=\! E_g \!+\! \frac{[-i\hbar\vec\nabla_{\!e} \!-\! (eBz_e/c) {\bf e}_y]^2}{2m_e} \!+\! W_{\!e}(z_e) \!-\! eEz_e +\\
 \frac{[-i\hbar\vec\nabla_{\!h} \!+\! (eBz_h/c){\bf e}_y]^2}{2m_h} \!+\! W_{\!h}(z_h) + eEz_h - \frac{e^2}{\varepsilon|\vec{r}_e \!-\! \vec{r}_h|},
\end{multline}
describing the relative motion of the electron and hole inside an exciton. In (\ref{Hrerh}), $m_{e(h)}$ is the electron (hole) effective mass, $e$ their charge modulus, $W_{\!e(h)}(z)$ denotes the QW potential for the electron (hole)(here assumed to have the shape of square wells of the width $L$), and the vector potential describing the magnetic field is chosen in the shape ${\bf A}_B(z)=-Bz{\bf e}_y$. As already noted, the Hamiltonians of the type (\ref{Hrerh}) do not conserve the exciton centre-of-mass (c.~m.) momentum. Instead, the operator of `magnetic momentum' now commutes with the Hamiltonian~\cite{gorkov,ruvinskii_pla,prb062015323}: in the chosen gauge for ${\bf A}_B$, it is given by $\hat{\vec{P}} = -i\hbar\vec{\nabla}_{\rm c.m.} +(e/c)B(z_e-z_h){\bf e}_y$. We recognize, however, that the extra term in $\hat{\vec{P}}$ is dependent only on the $z$-coordinates (while directed along the $y$-axis), hence upon separating the motion in the QW plane in~(\ref{Hrerh}), the in-plane momentum of the exciton c.~m. ${\bf p}$ becomes a good quantum number.

To calculate $\phi_{\bf p}$ and $E_{\bf p}$ we employ the variational method, similar to the one used in Ref.~\cite{prb075035303} for bulk GaAs excitons. In particular, we separate in $\phi_{\bf p}$ the plane wave $\exp\{i{\bf p \!\cdot\! r}/\hbar\}$ of the c.~m. motion, and for each ${\bf p}$ we minimize the functional
\begin{equation}\label{<phi|H0|phi>}
F[\phi_{\bf p}(\vec{r}_e,\vec{r}_h)]\equiv
F_{\bf p}(k_x,k_y;\lambda) \!=\!
\!\int\!\! d\vec{r}_e d\vec{r}_h\phi^*_{\bf p} \hat{H}_0 \phi_{\bf p}
\end{equation}
over the variational ansatz
\begin{equation}\label{phi_trial}
\phi_{\bf p} \!=\! \frac{e^{i{\bf p\cdot r}/\hbar}}{\sqrt{S}}
e^{i{\bf k}\cdot\boldsymbol{\varrho}/\hbar}\frac{2\lambda}{\sqrt{2\pi}}
e^{-\lambda\varrho}f_e(z_e,p_y,k_y)f_h(z_h,p_y,k_y),
\end{equation}
where ${\bf r}=(m_e{\bf r}_e+m_h{\bf r}_h)/m_{\rm ex}$ is the exciton c.~m. position and $\boldsymbol{\varrho}={\bf r}_e-{\bf r}_h$ is the relative in-plane coordinate of the electron and hole, while ${\bf k}=\{k_x,k_y\}$ and $\lambda$ are the variation parameters. The functions $f_{e(h)}(z,p_y,k_y)$ correspond to the ground state of the Schr\"{o}dinger equation for a single electron (hole)
\begin{multline}\label{Schred}
\!\!\left\{\! -\frac{\hbar^2\partial^2_{zz}}{2m_{e(h)}} \!+\!
\frac{e^2B^2z^2}{2m_{e(h)}c^2} \!+\! \left[\mp e\!\left(\!E +\frac{Bp_y}{m_{\rm ex}c}\!\right)\! -\frac{eBk_y}{m_{e(h)}c}\right]\!z \right. \\ \biggl.
+ W_{e(h)}(z) - \mathscr{E}_{e(h)}(p_y,k_y)\!\biggr\} f_{e(h)}(z,p_y,k_y)=0,
\end{multline}
with the upper (lower) sign corresponding to $e$ ($h$), eigenenergies denoted as $\mathscr{E}_{e(h)}$, and the normalization condition $\int_{-\infty}^{\infty}|f_{e(h)}|^2dz=1$. We note that the specific choice of gauge for ${\bf A}_B$ resulted in the one-dimensional equation with respect to $z$, whereas the motion in $(x,y)$--plane according to (\ref{phi_trial}) is given by plane-wave factors.

The Eq.~(\ref{Schred}) solution details are provided in Appendix~\ref{AppMin}. The electron and hole wavefunctions $f_{e,h}(z)$ are plotted in Fig.~\ref{fig1}{\bf b} for the QW width $L=30$~nm at zero magnetic field, for $E=5$~kV/cm. The dipole length calculated as $d/e = \int(z_e-z_h)|f_e(z_e)f_h(z_h)|^2dz_edz_h$ is plotted dependent on the electric field in Fig.~\ref{fig1}{\bf c} for $L=20$ and 30~nm. In the same panel, we plot the electron-hole overlap integral (in transverse direction) $I_z=\int f_e(z)f_h(z)dz$, as well as the Rabi splitting (\ref{OmegaRabi}), normalized to their values at zero electric field. One sees that for the 30-nm QW, the dipole length of 9~nm is achieved already at $E=5$~kV/cm which corresponds to the drop of the Rabi splitting to 56\% of its zero-field value (marked by the black arrow). For $L=20$~nm, the 9-nm dipole is only achieved at a much stronger field $E\approx34.3$~kV/cm which results in the decrease of $\hbar\Omega_0$ by three quarters. It is due to this reason that we argue that the wide QW should be considered in order to maintain the oscillator strength and provide conditions for dipolariton formation and BEC.

For the wavefunction of the exciton in cross electric and magnetic fields,  the minimization problem yields:
\begin{equation}\label{phip}
\phi_{\bf p}(\vec{r}_e,\vec{r}_h) \!=\! \frac{e^{i{\bf p\cdot r}/\hbar}}{\sqrt{S}}e^{iq_0\varrho_y/\hbar} \frac{2\lambda_0}{\sqrt{2\pi}} e^{-\lambda_0\varrho}f_e(z_e)f_h(z_h),
\end{equation}
with $q_0 = (eB\mu_{eh}/c)(\bar{z}_e/m_e + \bar{z}_h/m_h)$ [see Eq.~(\ref{q0})], where $\mu_{eh}$ is the electron-hole reduced mass, and the value of $\lambda_0$ found from the condition maximizing the exciton binding energy
\begin{equation}\label{Eb}
E_b(\lambda) \!=\! -\frac{\hbar^2\lambda^2}{2\mu_{eh}} \!+\!
\frac{e^2}{\varepsilon}\frac{2\lambda^2}{\pi}
\!\!\!\int\limits_{-\infty}^{\infty}\!\!\!dz_edz_hd\boldsymbol{\varrho}\,
\frac{e^{-2\lambda\varrho}|f_{\!e}(z_e)f_{\!h}(z_h)|^2}{\sqrt{\!\varrho^2 \!+\! (z_e \!-\! z_h)^2}}.
\end{equation}
Here we took into account that for positive $E$ and $B$, the exciton dipole moment is positive ($d>0$), and that the variational parameters at the minimum do not depend on momentum ${\bf p}$, being equal $k_x=0$, $k_y=q_0$, $\lambda=\lambda_0$.
The corresponding exciton dispersion has the form
\begin{equation}\label{Ep}
E_{\bf p}=F_{\bf p}(0,q_0,\lambda_0)=E_G+\frac{({\bf p-p}_0)^2}{2m_{\rm ex}},
\end{equation}
where ${\bf p}_0=Bd{\bf e}_y/c$ is the displacement momentum, and
\begin{equation}\label{EG}
E_G=E_g + \mathscr{E}_e + \mathscr{E}_h - \frac{B^2d^2}{2m_{\rm ex}c^2} - \frac{q_0^2}{2\mu_{eh}} - E_b(\lambda_0)
\end{equation}
is the renormalized (in crossed fields) exciton gap. We note that even though we deal everywhere with the exciton in-plane c.~m. momentum ${\bf p}$, the shift of the exciton dispersion (\ref{Ep}) by ${\bf p}_0\perp{\bf B}$ happens because the magnetic momentum $\hat{\vec{P}}$ is the actual integral of motion.

The minimisation problem is solved self-consistently for each fields strength combination $(E,B)$ in consideration, providing the exciton spectrum (\ref{Ep}) and the wavefunctions that are needed to define the Rabi splitting (\ref{OmegaRabi}) and the exciton field operator (\ref{Qp}) that are to be used in the Hamiltonian (\ref{Hexph}). An example of the shifted exciton dispersion versus $p_y$ (the direction of ${\bf p}_0$) is plotted in Fig.~\ref{fig2}{\bf a} by the dashed line for $E=4.2$~kV/cm and $B=3$~T. For these fields strengths, the dipole length $d/e= 7$~nm. We note that the presence of the in-plane magnetic field results in the shortening of the exciton dipole due to the diamagnetic terms $\sim B^2$ in Eq.~(\ref{Schred}).

\subsection{Dipolariton dispersion in crossed fields}

\begin{figure}[b]
  \includegraphics[width=\linewidth]{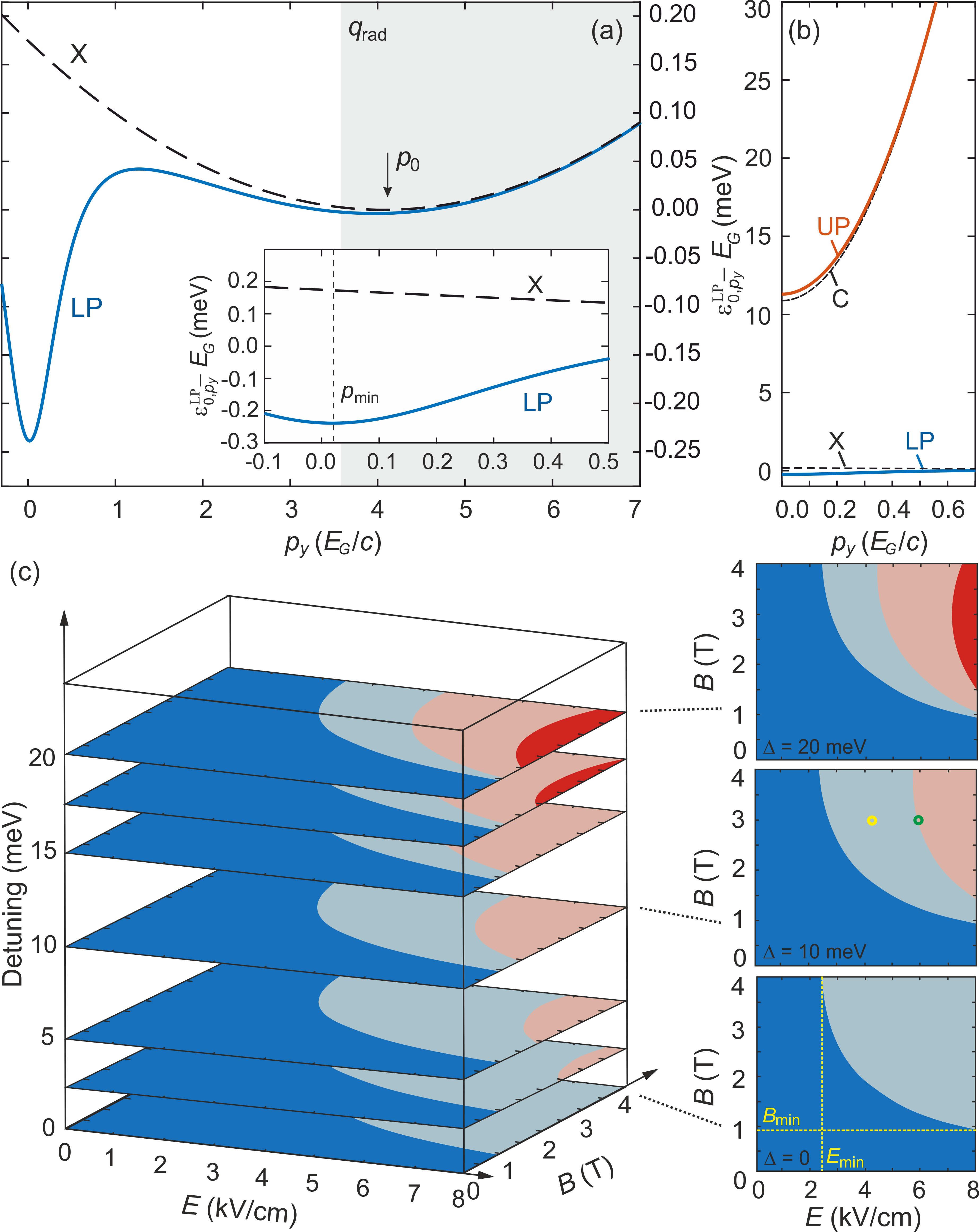}
  \caption{(a) Single-particle dispersions of excitons [according to Eq.~(\ref{Ep}), the dashed lines marked `X'] and lower polaritons [Eq.~(\ref{e0p}), the blue solid lines marked `LP'] versus $p_y$ at $p_x=0$ in a wide QW ($L=30$~nm) in transverse electric field $E=4.2$~kV/cm and in-plane magnetic field $B=3$~T, with the photon-exciton detuning $\Delta = 10$~meV. The grey-shaded region marks momenta lying outside the lightcone of the material (for GaAs, $q_{\rm rad}\approx 3.53 E_G/c$). The inset shows the magnified view of the near-zero region, revealing the polariton minimum shifted from ${\bf p}=0$. (b) Full view of the two polariton branches, where `C' denotes the cavity photon dispersion $\hbar\omega_{\bf p}$ and `UP' the upper-polariton dispersion.
  (c) Diagram of existence of the two minima in the LP dispersion dependent on $B$, $E$, and the detuning $\Delta$. Dark blue: only one (polariton) minimum; light blue: two minima with the polariton minimum deeper than the exciton minimum; pink: two minima with the exciton minimum deeper; red: only one (exciton) minimum present. The yellow mark indicates the parameters of the panels (a--b). The green mark indicates the parameters of Fig.~\ref{fig3b}. For all panels, the Rabi splitting in the absence of the fields $\hbar\Omega_0=6$~meV.}
  \label{fig2}
\end{figure}

After obtaining the single-particle exciton dispersion (\ref{Ep}) and the wavefunction (\ref{phip}) in the presence of electric and magnetic fields, we can proceed with diagonalizing the quadratic (kinetic) term in the exciton-photon Hamiltonian (\ref{Hexph}) and dressing of the bare exciton interaction. As a result, the Hamiltonian of the system in the dressed shape takes the form
\begin{equation}\label{Hmomentum}
\hat{H}=\sum\limits_{\bf p}\bigl(\varepsilon^{\rm LP}_{\bf p}
\hat{a}_{\bf p}^{\dagger}\hat{a}_{\bf p} + \varepsilon^{\rm UP}_{\bf p}\hat{b}_{\bf p}^\dag\hat{b}_{\bf p}\bigr) + \hat U_{\rm ex},
\end{equation}
where $\hat{a}_{\bf p} \!=\! X_{\bf p}\hat{Q}_{\bf p} \!+\! \sqrt{1 \!-\! X_{\bf p}^2}\hat{c}_{\bf p}$, $\hat{b}_{\bf p} \!=\! -\sqrt{1 \!-\! X_{\bf p}^2}\hat{Q}_{\bf p} \!+\! X_{\bf p}\hat{c}_{\bf p}$ are the annihilation operators of the lower (LP) and upper (UP) polaritons, respectively,
\begin{equation}\label{e0p}
\varepsilon^{\rm LP}_{\bf p} \!=\! E_{\bf p} \!+\! \frac{1}{2}\!\left[\hbar\omega_{\bf p} \!-\! E_{\bf p} \!-\! \sqrt{(\hbar\omega_{\bf p} \!-\! E_{\bf p})^2 \!+\! (\hbar\Omega_{\bf p})^2}\right]
\end{equation}
and $\varepsilon^{\rm UP}_{\bf p}=E_{\bf p}+\hbar\omega_{\bf p}-\varepsilon^{\rm LP}_{\bf p}$ are their respective dispersions. The exciton Hopfield coefficient is given by
\begin{equation}\label{Xp}
X_{\bf p}^2=\frac{1}
{1 + \left[\Delta_{\bf p}/(\hbar\Omega_{\bf p}) -
\sqrt{\Delta_{\bf p}^2/(\hbar\Omega_{\bf p})^2 +1}\right]^2},
\end{equation}
where $\Delta_{\bf p}=\hbar\omega_{\bf p}-E_{\bf p}$ is the detuning between the cavity photon and the exciton dispersions at the in-plane momentum ${\bf p}$. The last term in (\ref{Hmomentum}) denotes the dressed exciton-exciton interaction (see Appendix~\ref{AppU}) expressed via the LP particle operators $\hat a_{\bf p}$ and $\hat a_{\bf p}^{\dagger}$.

We note that since the Rabi splitting $\hbar\Omega_{\bf p}$ (\ref{OmegaRabi}) depends on the applied external fields via the exciton wavefunction, the Hopfield coefficient (\ref{Xp}) becomes also dependent on $E$ and $B$. Furthermore, due to the displacement of the exciton dispersion $E_{\bf p}$ at $B\neq0$ with respect to the minimum of the photon dispersion $\hbar\omega_{\bf p}$, the LP dispersion given by Eq.~(\ref{e0p}) shows a competition of the two minima appearing due to hybridization. An example of such a dispersion is plotted versus $p_y$ (at $p_x=0$) in Fig.~\ref{fig2}{\bf a--b} for $E=4.2$~kV/cm, $B=3$~T and $\Delta_{{\bf p}=0}\equiv\Delta=10$~meV. One sees two pronounced minima, near ${\bf p}=0$ and near ${\bf p} = {\bf p}_0$, both slightly shifted from these respective values (see the inset of Fig.~\ref{fig2}). Since ${\bf p}_0$ is directed along ${\bf e}_y$, the obtained polariton dispersion is neither centrally symmetric nor even (with respect to momentum), but there is a symmetry with respect to $p_x$ inversion. It is worth noting that the scale of the vertical axis in Fig.~\ref{fig2}{\bf a} and the difference in depth of the two minima is of the order of fractions of meV, since the exciton dispersion compared to the photon one is flat ($m_{\rm ex}\gg m_{\rm ph}$, see Fig.~\ref{fig2}{\bf b}). However, the temperatures that we consider ($\sim1$~K) and the positive detunings $\Delta>\hbar\Omega_0$ provide long particle lifetimes, good thermalization, and narrow linewidth. Furthermore, in wide QWs in weak electric fields effects of disorder are suppressed~\cite{jetpl0830553,jetpl0840222,prb046010193}. Therefore, given the sample quality is high enough, the dipolariton dispersion reported in Fig.~\ref{fig2}{\bf a} should be observable.

To investigate the existence and the competition of the two minima in the LP dispersion, we plot a diagram in the parameter space $(B,E,\Delta)$ in Fig.~\ref{fig2}{\bf c} which shows the fields values at which the second minimum appears in the lower-polariton spectrum (light-blue region). One sees that there are minimal values of the fields strengths $E_{\rm min}$, $B_{\rm min}$ independent of $\Delta$ that are required to reach the regime where the LP dispersion starts to soften around ${\bf p}_0$. Still higher fields are required (here dependent on $\Delta$) to reach the regime when the ``exciton'' minimum starts to be deeper than the ``polariton'' one near ${\bf p}=0$ (pink region). Finally, at high enough detunings and electric field values, there is a regime when the dispersion features only one exciton minimum (red region), which corresponds to the loss of the exciton oscillator strength and the quenching of the Rabi coupling. Note that since $p_0 \!\sim\! Bd$, with the increase of the fields strengths the exciton minimum moves out of the radiative zone of the material (the grey-shaded area in Fig.~\ref{fig2}{\bf a}).

Since the characteristic energies of the system are of the order of 0.1--1~meV, while the photon-to-exciton energy detuning in our consideration has the order of 10~meV, such a system thermalizes during its lifetime~\cite{deng2006,prb104125301}. Therefore both thermal (Boltzmann) and zero-temperature (quantum) occupations of the UP branch are negligibly small, and it is justified to assume that the upper polaritons are absent, set $\hat{b}_{\bf p}=0$ in~(\ref{Hmomentum}), and express the exciton and photon operators as
\begin{equation}\label{Qca}
\hat{Q}_{\bf p}=X_{\bf p}\hat{a}_{\bf p}, \;\;\;\; \hat{c}_{\bf p}=\sqrt{1-X_{\bf p}^2}\hat{a}_{\bf p}
\end{equation}
while keeping the bosonic commutation relations $[\hat a_{\bf p},\hat a_{{\bf p}^\prime}]=0$ and $[\hat a_{\bf p},\hat a_{{\bf p}^\prime}^{\dagger}]=\delta_{\bf pp^\prime}$ valid.

\section{The Bogoliubov theory and phase transition} \label{B-approx}

Focusing on the BEC regime, it is important to note that due to the peculiarities of the Hamiltonian (\ref{Hmomentum}) and the competition of the two minima in the single-particle dispersion (\ref{e0p}), the macroscopic uniform equilibrium system acquires a new free parameter: the condensate momentum ${\bf K}$. Settling in either of the two minima at nonzero momenta, the condensate will be at rest with zero group velocity.

In order to explicitly separate the condensate momentum in the system Hamiltonian, we define the integral convolution
\begin{equation}\label{e0-ihnabla}
\varepsilon_{\rm LP}(-i\hbar\nabla)f({\bf r}) \!=\! \frac{1}{S} \sum\limits_{\boldsymbol{k}}
\varepsilon_{{\bf K} + \boldsymbol{k}}^{\rm LP}\!\!\int \!\! e^{i\boldsymbol{k}\cdot({\bf r} - {\bf r}^\prime\!)/\hbar}f({\bf r}^\prime)d{\bf r}^\prime
\end{equation}
and the position-dependent function
\begin{equation}\label{Xr}
X({\bf r})=\frac1S\sum\limits_{\boldsymbol{k}}X_{{\bf K}+\boldsymbol{k}}e^{i\boldsymbol{k}\cdot{\bf r}/\hbar}
\end{equation}
which allows us to rewrite the exciton and polariton field operators [see (\ref{Qca})], respectively, as
\begin{align}
\hat{Q}({\bf r}) &=\!\int\!\! X({\bf r} \!-\! {\bf r}^\prime)\hat{\Psi}({\bf r}^\prime) d{\bf r}^\prime, \label{Qr}\\
\hat{\Psi}({\bf r}) &=\! \frac{1}{\sqrt S} \!\sum\limits_{\boldsymbol{k}} \!\hat{a}_{{\bf K}+\boldsymbol{k}}e^{i\boldsymbol{k}\cdot{\bf r}/\hbar}. \label{Psir}
\end{align}

Then after transformations the dressed Hamiltonian of the system (\ref{Hmomentum}) takes its final shape with the explicit dependence on ${\bf K}$ via the Eqs.~(\ref{e0-ihnabla})--(\ref{Psir}):
\begin{multline}\label{H}
\hat H= \int\hat\Psi^{\dagger}({\bf r})\varepsilon_{\rm LP}(-i\hbar\nabla)
\hat\Psi({\bf r})d{\bf r} + \!\int\! \epsilon_0[\hat Q^{\dagger}({\bf r})\hat Q({\bf r})]d{\bf r}  \\
+ \frac{1}{2}\!\int[U_0({\bf r\!-\!s}) \!-\! g_0\delta({\bf r \!-\! s})] \hat{Q}^\dag({\bf r})\hat{Q}^\dag({\bf s})\hat{Q}({\bf s})\hat{Q}({\bf r})d{\bf r}d{\bf s},
\end{multline}
where $\epsilon_0(n_{\rm ex})$ is the part of the free energy per unit area responsible for exciton-exciton interaction (see Appendix~\ref{AppU} and Ref.~\cite{e0NormalOrdering} for details), $U_0$ is the pair interaction potential (\ref{U0rho}).

Assuming the presence of macroscopic BEC, we will build the Bogoliubov theory~\cite{Bogoliubov} for lower polaritons in crossed fields.
Given the losses are small (for positive detunings), we consider the system as macroscopic, spatially uniform, and in thermal equilibrium.
In the case of macroscopic coherence, the operator of the total number of particles $\hat N\equiv\int\hat\Psi^{\dagger}({\bf r})\hat\Psi({\bf r})d{\bf r}\approx
N$ and the operator of condensate mode $\hat a_{\bf K}\approx\sqrt{N_0}$
are numbers, hence we express the polariton field operator and the total polariton density via the condensate density $n_0\equiv N_0/S$ ($N_0\equiv\langle\hat{a}_{\bf K}^{\dagger}\hat{a}_{\bf K}\rangle$ is the number of particles in the BEC):
\begin{equation}\label{Psirn0}
\hat{\Psi}({\bf r})=\sqrt{n_0}+\frac1{\sqrt S}\sum\limits_{\boldsymbol{k}\ne0} \hat{a}_{{\bf K} + \boldsymbol{k}}e^{i\boldsymbol{k}\cdot{\bf r}/\hbar},
\end{equation}
\begin{equation}\label{nn0}
n\equiv\frac{N}{S} = n_0 + \frac{1}{2S}\!\sum\limits_{\boldsymbol{k}\ne0}
(\hat{a}_{{\bf K}+\boldsymbol{k}}^{\dagger}\hat{a}_{{\bf K}+\boldsymbol{k}} + \hat{a}_{{\bf K}-\boldsymbol{k}}^{\dagger}\hat{a}_{{\bf K}-\boldsymbol{k}}).
\end{equation}

We assume that the condensate depletion is small and expand the Hamiltonian (\ref{H}) in the zero and first orders in $(n-n_0)/n_0$ [i.e. in the zero and second orders with respect to the non-condensate operators $\hat{a}_{{\bf K}+\boldsymbol{k}}$ ($\boldsymbol{k}\ne0$)]. Expressing the condensate density $n_0$ via the total density $n$ using Eq.~(\ref{nn0}), we neglect in $\hat{H}$ the cubic and quartic terms with respect to non-condensate operators. In particular, we substitute the field operator (\ref{Psirn0}) into the kinetic term in~(\ref{H}) and into the exciton field~(\ref{Qr}), which is then substituted into interaction part of the Hamiltonian~(\ref{H}) [see (\ref{U1B})].
Expanding the function $\epsilon_0(n_{\rm ex})$ in Taylor series around the point $n_{\rm ex}=nX_{\bf K}^2$, we obtain
\begin{multline}\label{T1}
\frac{\hat H}{S} = \varepsilon_{\bf K}^{\rm LP}n + \frac{1}{2S} \! \sum_{\boldsymbol{k}\ne0}\!\left[(\varepsilon_{{\bf K}+\boldsymbol{k}}^{\rm LP}-\varepsilon_{\bf K}^{\rm LP})
\hat{a}_{{\bf K}+\boldsymbol{k}}^{\dagger}\hat{a}_{{\bf K}+\boldsymbol{k}} \right. \\
\left.  \qquad\qquad\qquad\qquad\quad +\, (\varepsilon_{{\bf K}-\boldsymbol{k}}^{\rm LP}-\varepsilon_{\bf K}^{\rm LP}) \hat{a}_{{\bf K}-\boldsymbol{k}}^{\dagger}\hat{a}_{{\bf K}-\boldsymbol{k}}\right] \\
+  \epsilon_0(nX_{\bf K}^2) + \frac{\epsilon_0^\prime(nX_{\bf K}^2)}{2S}\! \sum_{\boldsymbol{k}\ne0} \! \left[(X_{{\bf K}+\boldsymbol{k}}^2-X_{\bf K}^2)\hat{a}_{{\bf K}+\boldsymbol{k}}^{\dagger} \hat{a}_{{\bf K}+\boldsymbol{k}} \right. \\
\qquad\qquad\qquad\qquad\quad\left. +
(X_{{\bf K}-\boldsymbol{k}}^2-X_{\bf K}^2)\hat{a}_{{\bf K}-\boldsymbol{k}}^{\dagger}\hat{a}_{{\bf K}-\boldsymbol{k}}\right] \\
+ \frac{nX_{\bf K}^2}{2S}\!\sum_{\boldsymbol{k}\ne0}\!U(\boldsymbol{k})\!\left[
X_{{\bf K}+\boldsymbol{k}}^2\hat a_{{\bf K}+\boldsymbol{k}}^{\dagger}\hat a_{{\bf K}+\boldsymbol{k}} \!+\! X_{{\bf K}-\boldsymbol{k}}^2\hat a_{{\bf K}-\boldsymbol{k}}^{\dagger}\hat a_{{\bf K}-\boldsymbol{k}}
\right. \\
\left. + X_{{\bf K}+\boldsymbol{k}}X_{{\bf K}-\boldsymbol{k}}(\hat{a}_{{\bf K}+\boldsymbol{k}}^{\dagger}\hat{a}_{{\bf K}-\boldsymbol{k}}^{\dagger} + \hat{a}_{{\bf K}+\boldsymbol{k}}\hat{a}_{{\bf K}-\boldsymbol{k}})\right]
\end{multline}
with
$U(\boldsymbol{k})=\epsilon_0^{\prime\prime}(nX_{\bf K}^2)+U_0(\boldsymbol{k})-g_0$
being the dressed exciton-exciton interaction, and $\epsilon_0^\prime$, $\epsilon_0^{\prime\prime}$ denoting the first and second derivatives of the function $\epsilon_0$ with respect to its argument. In derivation of Eq.~(\ref{T1}), we took into account that the Hopfield coefficients $X_{\bf p}$ and interaction $U({\bf p})=U({\bf -p})$ are real-valued.

Substituting the obtained expression into the free energy, one finds to the leading (zero) order the momentum ${\bf K}_0$ of the condensate at rest. Assuming for simplicity the periodic boundary conditions ${\bf K} =(2\pi\hbar/L){\bf l}$ (here $L=\sqrt{S}$ is the system in-plane size) with ${\bf l}\in\mathbb{Z}^2$, we define ${\bf K}_0$ from the minimization of the free energy per unit area
\begin{equation}\label{F=min}
\mathscr{F}=\min_{\bf K}\langle\hat{H}/S\rangle
\end{equation}
over all values of ${\bf K}$ (i.e. over all values of the integer-valued 2D vector ${\bf l}$).
Namely, to find ${\bf K}_0$ we minimize the function
\begin{equation}\label{F(K)=min}
\mathscr{F}({\bf K})=\varepsilon_{\bf K}^{\rm LP}n + \epsilon_0(nX_{\bf K}^2)
\end{equation}
over all values of ${\bf K}$. The minima of $\mathscr{F}$ differ from the minima of the bare particle dispersion due to the extended range of interactions that are brought in the system by the dipolar excitons.
Derivating Eq.~(\ref{F(K)=min}) with respect to the full density $n$, we find the chemical potential of the system of lower polaritons at a given ${\bf K}$:
\begin{equation}\label{mu}
\mu\equiv\frac{\partial\mathscr{F}}{\partial n} = \varepsilon_{\bf K}^{\rm LP} + \epsilon_0^\prime(nX_{\bf K}^2)X_{\bf K}^2.
\end{equation}

\begin{figure}[t!]
\includegraphics[width=\linewidth]{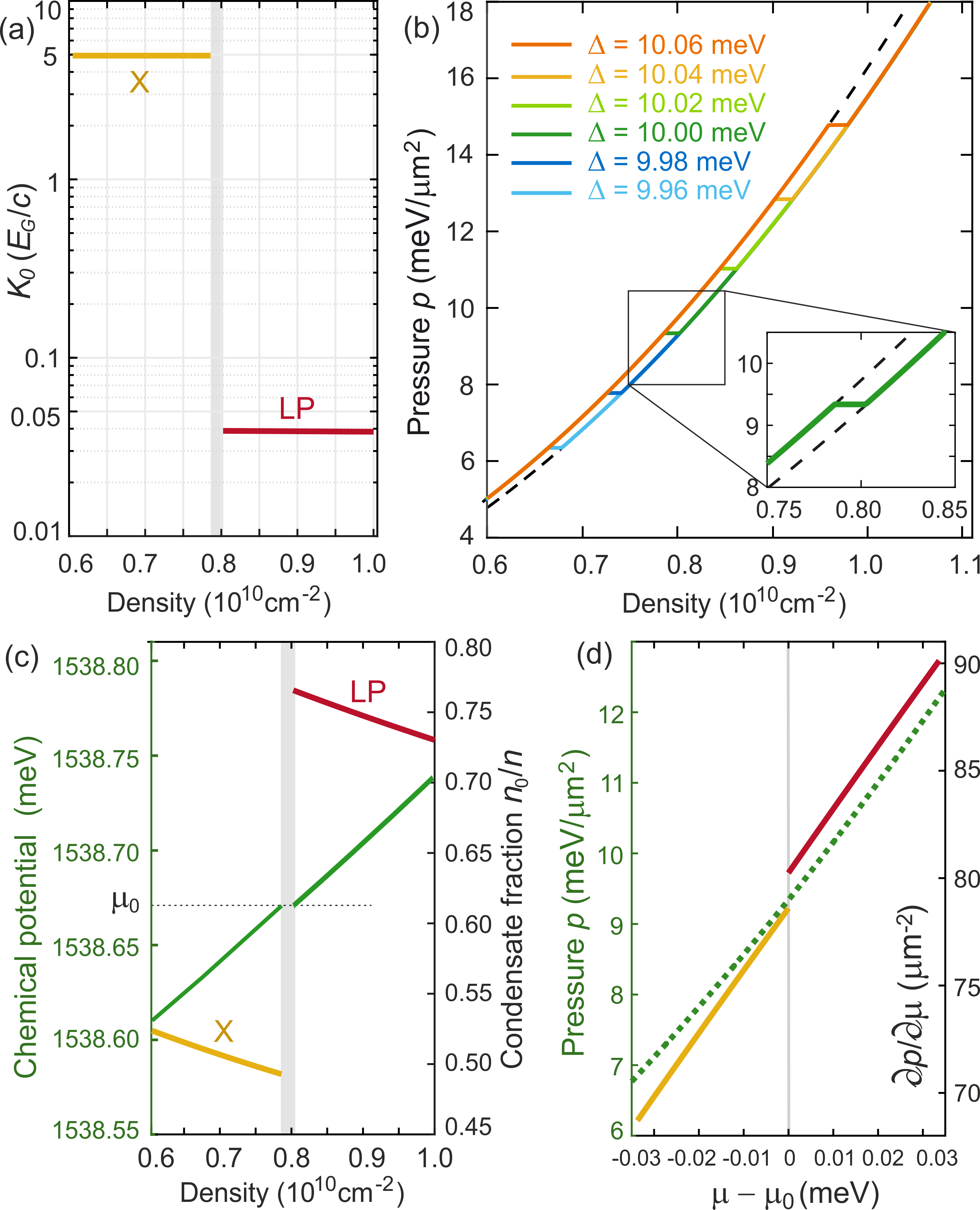}
\caption{First-order phase transition from the exciton to the lower-polariton BEC. (a) Condensate momentum ${\bf K}={\bf K}_0$ (absolute value) across the transition from the exciton--BEC regime (the yellow line) to  the polariton--BEC regime (the red line). At $n = 0.78\times 10^{10}$~cm$^{-2}$ on the exciton side of the transition, $K_0 = 7595.38$~meV$/c$, while at $n = 0.81\times 10^{10}$~cm$^{-2}$, i.e. in the polariton regime, $K_0 = 59.85$~meV$/c$. (b) Pressure against total density across the transition, for different detunings (as given in the legend). The dashed lines indicate the pressure when in the exciton-BEC and polariton-BEC regimes, while the region between them corresponds to coexistence of two phases.  The inset shows a magnified view of the dependence $p(n)$ in the vicinity of transition for $\Delta = 10.0$~meV. (c) Chemical potential $\mu$ (left axis) according to Eq.~(\ref{mu}) and condensate fraction $n_0/n$ (right axis) across the transition. $\mu_0$ indicates the chemical potential of the system at the transition. (d) Pressure and its derivative with respect to chemical potential versus $\mu-\mu_0$. While the pressure is continuous at the transition, its derivative exhibits a pronounced discontinuity, indicating the first-order type of transition. For all panels, $B=3$~T, $E=5.9$~kV/cm, $d/e = 8.5$~nm, detuning $\Delta = 10.0$~meV (except (b)), $\hbar\Omega_0=6$~meV at $E=B=0$ and 3.5~meV in the applied $E$ and $B$. The renormalized exciton gap $E_G = 1538.46$~meV. The grey-shaded areas in (a),(d) indicate the transition region $n\in[0.7862,0.8027]$ at $\Delta = 10.0$~meV.
}
\label{fig3}
\end{figure}

The minimization procedure allows us to find the condensate momentum ${\bf K}_0$ for each $E$ and $B$ depending on the total density $n$ and detuning $\Delta$. In Fig.~\ref{fig3}{\bf a}, we plot an example of such a dependence for $B=3$~T, $E=5.9$~kV/cm, $\Delta=10$~meV (at $\hbar\Omega_0=6$~meV). 
Upon changing $n$ at a fixed detuning, we evidence that
the resting (${\bf K=K}_0$) superfluid Bose-condensed system of excitons, 
due to the energy considerations---according to Eq.~(\ref{F=min})---undergoes a transition from its exciton minimum of the free energy to the polariton one (for an exemplary dependence of the free energy (\ref{F(K)=min}) on ${\bf K}$ and details of the transition produced by changing the detuning at a fixed density, see Appendix~\ref{AppDet}). 
The condensate magnetic momentum $K_0$ upon transiting to the regime of the lower polaritons BEC changes abruptly by two orders of magnitude: from  4.937 $E_G/c$ below $n=0.7862\times10^{10}$~cm$^{-2}$ to 0.0389 $E_G/c$ at $n$ higher than $0.8027\times10^{10}$~cm$^{-2}$. We note that the value of the condensate momentum $K_0$ in the exciton-BEC regime at $\varepsilon=12.5$ (for GaAs) exceeds the lightcone radius $q_{\rm rad} = E_G\sqrt{\varepsilon}/c = 3.53E_G/c$. This means that the exciton condensate is optically dark~\cite{prb062001548,prb062015323}, so that it decays mostly non-radiatively, featuring very long (on the $\mu$s scale~\cite{jetpl0840222}) lifetimes (see discussion in Sec.~\ref{B-opt}).

In order to investigate the nature of this transition, we calculate the two-dimensional pressure as $p(\mu)=-[\mathscr{F}_{\rm min} - \mu n(\mu)]$ which is plotted against the total density in Fig.~\ref{fig3}{\bf b} for different values of $\Delta$. The chemical potential according to (\ref{mu}) is plotted for the same values of $n$ in Fig.~\ref{fig3}{\bf c}. One sees that for each detuning, there exists a narrow range of densities corresponding to the coexistense of the two BEC phases where both the pressure and chemical potential stay constant. The pressure dependence on $\mu$ stays continuous, as shown in Fig.~\ref{fig3}{\bf d}. At the same time, the derivative $\partial p/\partial\mu$ displays a jump across the exciton-BEC---polariton-BEC transition, thus indicating that this is a first-order phase transition, tunable by means of total population, detuning, or electric field (see Appendix~\ref{AppDet}).

Furthermore, we study the excitation spectrum of such a system, its condensate population, and their change across the considered transition. First, we bring the Hamiltonian (\ref{T1}) to the traditional Bogoliubov shape convenient for diagonalization. To shorten the derivations, we introduce the following notation:
\begin{multline}\label{T(p)}
T(\boldsymbol{k}) = \varepsilon_{{\bf K}+\boldsymbol{k}}^{\rm LP} + \epsilon_0^\prime(nX_{\bf K}^2)X_{{\bf K}+\boldsymbol{k}}^2 \\
+ U(\boldsymbol{k})nX_{\bf K}^2X_{{\bf K}+\boldsymbol{k}}(X_{{\bf K}+\boldsymbol{k}}-X_{{\bf K}-\boldsymbol{k}}),
\end{multline}
which allows to define the ${\bf K}$-dependent symmetric kinetic function of momentum~\cite{footnote2}
\begin{equation}\label{Tp}
\mathcal{T}_{\boldsymbol{k}} \!=\! \frac{T(\boldsymbol{k}) \!-\! 2T(0) \!+\! T(-\boldsymbol{k})}{2},\qquad \mathcal{T}_{\boldsymbol{k}}=\mathcal{T}_{-\boldsymbol{k}},
\end{equation}
where $\boldsymbol{k} = {\bf p}-{\bf K}$. Similarly, we introduce the symmetric potential function
\begin{equation}\label{Up}
\mathcal{U}_{\boldsymbol{k}}=
U(\boldsymbol{k})X_{\bf K}^2X_{{\bf K}+\boldsymbol{k}}X_{{\bf K}-\boldsymbol{k}}n, \qquad \mathcal{U}_{\boldsymbol{k}}=\mathcal{U}_{-\boldsymbol{k}},
\end{equation}
and the asymmetric function
\begin{equation}\label{UpAp}
\mathcal{A}_{\boldsymbol{k}} \!=\! \frac{T(\boldsymbol{k}) \!-\! T(-\boldsymbol{k})}2,\qquad \mathcal{A}_{\boldsymbol{k}}=-\mathcal{A}_{-\boldsymbol{k}}.
\end{equation}
As a result, after transformations the dressed Hamiltonian (\ref{H}) is finally rewritten as
\begin{multline}\label{H1}
\hat{H} - \mu\hat{N} \!=\! \frac{1}{2}\!\sum\limits_{\boldsymbol{k}\ne0} \!
\left[(\mathcal{T}_{\boldsymbol{k}} \!+ \mathcal{U}_{\boldsymbol{k}})
(\hat{a}_{{\bf K}+\boldsymbol{k}}^{\dagger}\hat{a}_{{\bf K}+\boldsymbol{k}} \!+ \hat{a}_{{\bf K}\!-\!\boldsymbol{k}}^{\dagger}\hat{a}_{{\bf K}\!-\!\boldsymbol{k}})\right. \\
+ \mathcal{U}_{\boldsymbol{k}} (\hat{a}_{{\bf K} +\boldsymbol{k}}^{\dagger}\hat{a}_{{\bf K}-\boldsymbol{k}}^{\dagger} + \hat{a}_{{\bf K}+\boldsymbol{k}}\hat{a}_{{\bf K}-\boldsymbol{k}}) \qquad \\
\left. + \mathcal{A}_{\boldsymbol{k}} (\hat{a}_{{\bf K}+\boldsymbol{k}}^{\dagger}\hat{a}_{{\bf K}+\boldsymbol{k}} -\hat{a}_{{\bf K}-\boldsymbol{k}}^{\dagger}\hat{a}_{{\bf K}-\boldsymbol{k}})\right] + \text{const}.
\end{multline}
The non-condensate part can be diagonalized using the Bogoliubov transformation
\begin{equation}\label{BT}
\hat{a}_{{\bf K}+\boldsymbol{k}}=u_{\boldsymbol{k}}\hat{\alpha}_{\boldsymbol{k}} - v_{\boldsymbol{k}}\hat{\alpha}_{-\boldsymbol{k}}^{\dagger},
\end{equation}
where $\hat{\alpha}_{\boldsymbol{k}}$ is the annihilation operator of the Bogoliubov excitation with momentum $\boldsymbol{k}\ne0$ above the mode ${\bf K}$, and the Bogoliubov amplitudes are given by
\begin{equation}\label{up2vp2ep}
u_{\boldsymbol{k}}^2,v_{\boldsymbol{k}}^2 \!=\! \frac{1}{2}\!\left(\!\sqrt{1 + \frac{\mathcal{U}_{\boldsymbol{k}}^2}{\mathcal{E}_{\boldsymbol{k}}^2}}\pm 1 \!\right)\!\!,\,\,\, \mathcal{E}_{\boldsymbol{k}}\!\equiv\! \sqrt{\mathcal{T}_{\boldsymbol{k}}(\mathcal{T}_{\boldsymbol{k}}+2\, \mathcal{U}_{\boldsymbol{k}})}.
\end{equation}

After the diagonalization, the Hamiltonian~(\ref{H}) takes the form
\begin{equation}\label{H=}
\hat{H}-\mu\hat{N}={\rm const}+\sum\limits_{\boldsymbol{k}\ne0}
\varepsilon_{\boldsymbol{k}}\hat\alpha_{\boldsymbol{k}}^{\dagger} \hat\alpha_{\boldsymbol{k}}, \quad \varepsilon_{\boldsymbol{k}} \equiv\mathcal{E}_{\boldsymbol{k}}+\mathcal{A}_{\boldsymbol{k}},
\end{equation}
where $\varepsilon_{\boldsymbol{k}}$ is the Bogoliubov spectrum of excitations, with the stability conditions
\begin{equation}\label{3stab}
\varepsilon_{\boldsymbol{k}}>0,\quad
\mathcal{T}_{\boldsymbol{k}}>0,\quad\mathcal{T}_{\boldsymbol{k}}+2\, \mathcal{U}_{\boldsymbol{k}}>0.
\end{equation}

\begin{figure}[t]
\includegraphics[width=\linewidth]{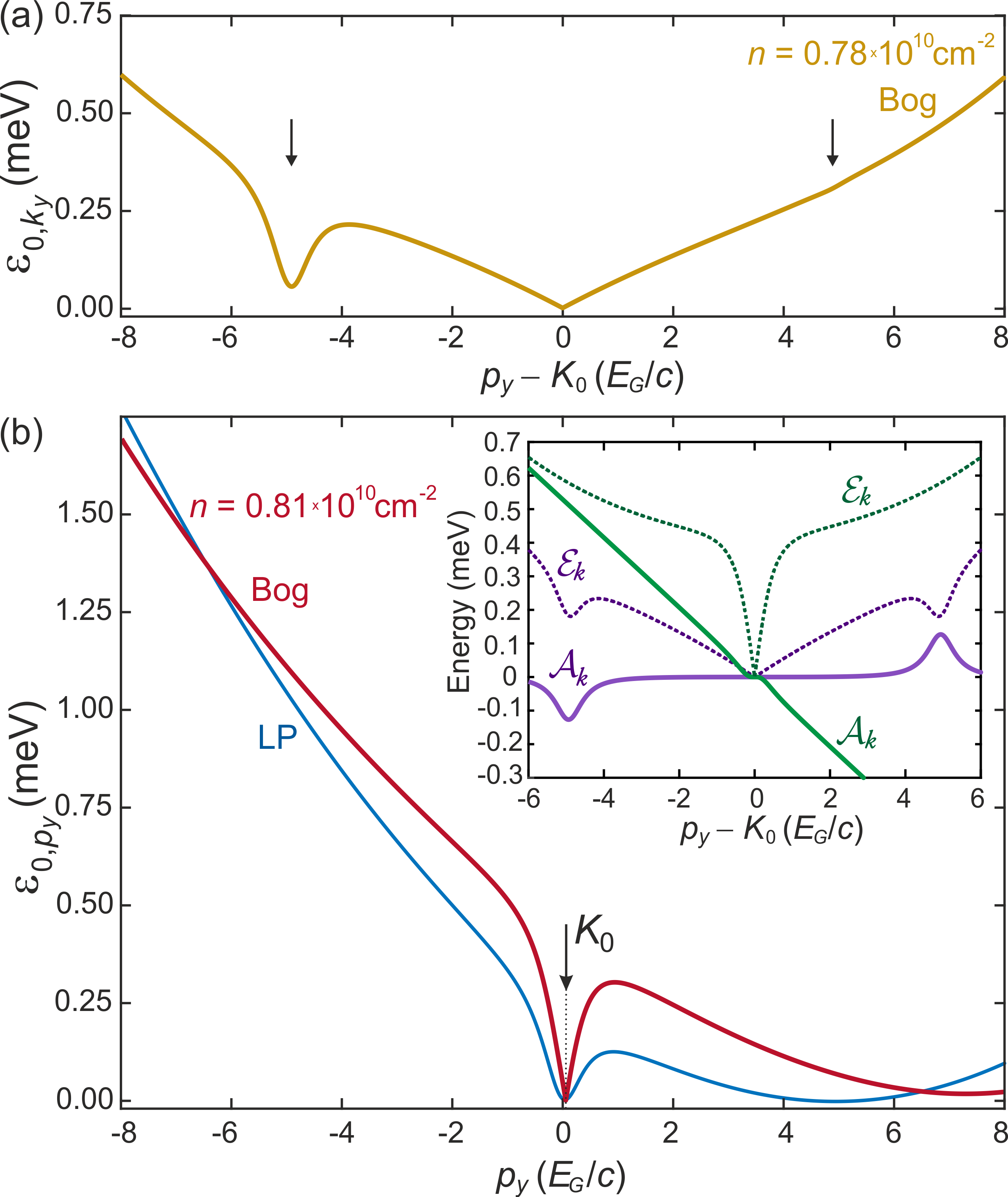}
\caption{The Bogoliubov spectrum of the system according to Eq.~(\ref{H=}) to the both sides of the transition, with the same parameters as in Fig.~\ref{fig3}. (a) The exciton-BEC excitation spectrum ($n=0.78\times10^{10}$~cm$^{-2}$). The two black arrows indicate the symmetric minima positions on the spectrum. Note that we build the Bogoliubov theory on top of the condensate with the momentum ${\bf K}_0$. (b) The red solid line corresponds to the excitation spectrum of the polariton BEC ($n=0.81\times10^{10}$~cm$^{-2}$). The thin blue line shows the single-particle LP dispersion $\varepsilon^{\rm LP}_{\bf p}-E_G$. Inset: The symmetric ($\mathcal{E}_{\boldsymbol{k}}$, dotted lines) and antisymmetric ($\mathcal{A}_{\boldsymbol{k}}$, solid lines) parts of the Bogoliubov spectrum (\ref{H=}) in the polariton regime (green) and exciton regime (purple). In (b), $K_0$ lies very close to $p_y=0$, and the horizontal axis represents the absolute value of in-plane momentum. In the inset and in (a), momentum is relative to ${\bf K}_0$.
}
\label{fig3b}
\end{figure}

We plot the excitation spectrum (\ref{H=}) for the detuning $\Delta=10$~meV  for the total densities of the polariton system $n$ corresponding to the two sides of the exciton-BEC---polariton-BEC transition in Fig.~\ref{fig3b}{\bf a} and {\bf b}.
In both regimes the Bogoliubov spectra of excitations $\varepsilon_{\boldsymbol{k}}$ are positive at all momenta (i.e. the system is stable). Furthermore, the Landau critical velocity for superfluidity $v_{\rm cr}=\min_{\boldsymbol{k}}(\varepsilon_{\boldsymbol{k}}/k)>0$ does not turn to zero. We note that $v_{\rm cr}$ is anisotropic and not even with respect to $k_y$, and stays much lower than the polariton sound velocity $c_s=\mathrm{lim}_{\boldsymbol{k}\to0}(\varepsilon_{\boldsymbol{k}}/k)$ (which can be attributed to the high-quality thermalization in the system).
In both BEC regimes, there is a pronounced roton-maxon effect in the Bogoliubov spectrum of excitations~$\varepsilon_{\boldsymbol{k}}$.
For clarity, the LP single-particle dispersion $\varepsilon^{\rm LP}_{\bf p}$ is displayed by the blue line in Fig.~\ref{fig3b}{\bf b}, revealing the two pronounced minima of approximately the same depth, in agreement with the minimum-competition diagram in Fig.~\ref{fig2}{\bf c} (see the green mark in the panel corresponding to $\Delta=10$~meV).
The Bogoliubov spectrum in the polariton-BEC regime (Fig.~\ref{fig3b}{\bf b}) has a wide softened region along the direction $k_y$, with the roton minimum depth (at given $E$, $B$, and $\Delta$) defined by the total particle density. Note that since the condensate momentum ${\bf K}_0$ in the polariton-BEC regime lies very close to ${\bf p}=0$ (see Fig.~\ref{fig3}{\bf a}), we plot the excitation spectrum in Fig.~\ref{fig3b}{\bf b} against the absolute momentum projection $p_y$.
The spectrum of excitations on top of the exciton BEC (Fig.~\ref{fig3b}{\bf a}), on the other hand, is plotted against $k_y=p_y-K_0$. It shows a narrow dip on the opposite side of the condensate (at ${\bf K}_0$), which occurs due to the presence of the polariton minimum in the LP dispersion. One also notes a symmetrically placed, extremely shallow minimum on the opposite side of $K_0$ (see the black arrows in Fig.~\ref{fig3b}{\bf a}). The appearance of these features is dictated by the asymmetry of the function $\mathcal{A}_{\boldsymbol{k}}$ [see (\ref{UpAp}) and the inset of Fig.~\ref{fig3b}{\bf b}]: being substituted in Eq.~(\ref{H=}), it results in the summation of the two functions for $k_y>0$ and their subtraction for $k_y<0$.

\section{Correlators}\label{sec_corr}

Knowing the field operator $\hat\Psi({\bf r})$ via the Eqs.~(\ref{Psirn0}),
(\ref{BT}) and the system Hamiltonian~(\ref{H=}), we can study various polariton and exciton correlations. In particular, we calculate the polariton occupation number
\begin{equation}\label{NpB}
N_{\boldsymbol{k}}\equiv\langle\hat a_{{\bf K}+\boldsymbol{k}}^{\dagger}\hat a_{{\bf K}+\boldsymbol{k}}\rangle = u_{\boldsymbol{k}}^2n_{\boldsymbol{k}}+v_{\boldsymbol{k}}^2(1+n_{-\boldsymbol{k}}),
\end{equation}
where $n_{\boldsymbol{k}}=\langle\hat\alpha_{\boldsymbol{k}}^{\dagger} \hat\alpha_{\boldsymbol{k}}\rangle = 1/(e^{\varepsilon_{\boldsymbol{k}}/T}-1)$ is the Bose distribution of the Bogoliubov excitations with the temperature $T$. In a similar fashion, we calculate the one-body density matrix of lower polaritons
\begin{multline}\label{g1rB}
g_1({\bf r})\equiv\langle\hat\Psi^{\dagger}({\bf r})\hat\Psi(0)\rangle=
n-\frac1S\sum\limits_{\boldsymbol{k}\ne0}(1-e^{-i{\boldsymbol{k}\cdot r}/\hbar})n_{\boldsymbol{k}}-
\\
-\frac1S\sum\limits_{\boldsymbol{k}\ne0}\left(1-\cos\frac{\boldsymbol{k}\!\cdot\! {\bf r}}{\hbar}\right) v_{\boldsymbol{k}}^2(1+2n_{\boldsymbol{k}})
\end{multline}
and their momentum-frequency distribution:
\begin{align}
N({\bf K}+\boldsymbol{k},\omega)& \equiv\!\!\int\limits_{-\infty}^{\infty}e^{i\omega t}
\langle\hat a_{{\bf K}+\boldsymbol{k}}^{\dagger}(0)\hat a_{{\bf K}+\boldsymbol{k}}(t)\rangle\frac{dt}{2\pi} \label{npo} \\
& = N_0\delta_{\boldsymbol{k}0}\delta(\omega) + \left[u_{\boldsymbol{k}}^2n_{\boldsymbol{k}}\delta\!\left(\omega - \frac{\varepsilon_{\boldsymbol{k}}}{\hbar}\right)\right.\nonumber \\
& \left. + v_{\boldsymbol{k}}^2(1 + n_{-\!\boldsymbol{k}}) \delta\!\left(\omega + \frac{\varepsilon_{-\!\boldsymbol{k}}}{\hbar}\right)\right]\!(1- \delta_{\boldsymbol{k}0}). \nonumber
\end{align}
We note that, as one of the main features of the polariton system in crossed fields, the Bose distribution of excitations $n_{\boldsymbol{k}}$, the occupation number $N_{\boldsymbol{k}}$ given by Eq.~(\ref{NpB}) and the momentum-frequency distribution $N({\bf K}+\boldsymbol{k},\omega)$ in~(\ref{npo}) are {\it not} even functions of momentum. Furthermore, the normal one-body density matrix $g_1({\bf r})$ is {\it complex-valued}. This occurrence is not an artefact of the developed theory, as all physical quantities calculated from Eq.~(\ref{g1rB}) are real: e.g., the optical interference signal in the Young experiment~\cite{jetpl0840329} for the central bright fringe contains $g_1({\bf r})+g_1({\bf -r})$.

To calculate the condensate density, we use the unification of the Bogoliubov approach with quantum hydrodynamics~\cite{prb104125301,Popov,pr0155000080,prl121235702,prb103094511}, which yields the expression
\begin{equation}\label{n0HDB}
n_0 = n_{\rm q} \exp\!\left[
-\frac1{N_{\rm q}}\sum\limits_{\boldsymbol{k}\ne0}v_{\boldsymbol{k}}^2 (1+2n_{\boldsymbol{k}})\right],
\end{equation}
where $N_{\rm q} = N-\sum_{\boldsymbol{k}\ne0}n_{\boldsymbol{k}}$ and $n_{\rm q} = N_{\rm q}/S$ are the quasicondensate particle number and density, respectively (in the theory of Berezinskii-Kosterlitz-Thouless (BKT) transition, true Bose condensation in 2D is replaced by the quasicondensate formation and the appearance of local superfluidity, see e.g.~\cite{prb104125301}). The result (\ref{n0HDB}) coincides with the prediction of the Bogoliubov theory $n_0 = n_{\rm q}-\sum_{\boldsymbol{k}\ne0} v_{\boldsymbol{k}}^2 (1+2n_{\boldsymbol{k}})/S$ [see (\ref{NpB})] up to the first order of the exponent expansion~\cite{prl121235702,prb104125301}.
Fixing the detuning $\Delta=10$~meV, we investigate the behaviour of the condensate fraction $n_0/n$ at $T=0$ according to (\ref{n0HDB}) across the transition (dependent on the density). The result is shown in Fig.~\ref{fig3}{\bf c}: when going from higher to smaller densities, the condensate fraction in the system drops from 0.77 in the polariton-BEC regime to 0.49 in the exciton-BEC regime, at fixed electric and magnetic fields $E=5.9$~kV/cm, $B=3$~T and the Rabi splitting $\hbar\Omega_0=6$~meV (at $E=B=0$). 
The drop of the condensate fraction while passing to the exciton regime to less than 50~\% indicates that the transition essentially changes the regime of correlations in the system, from weakly-correlated polariton BEC to the intermediately-correlated BEC of excitons.

\begin{figure}[t]
\includegraphics[width=\linewidth]{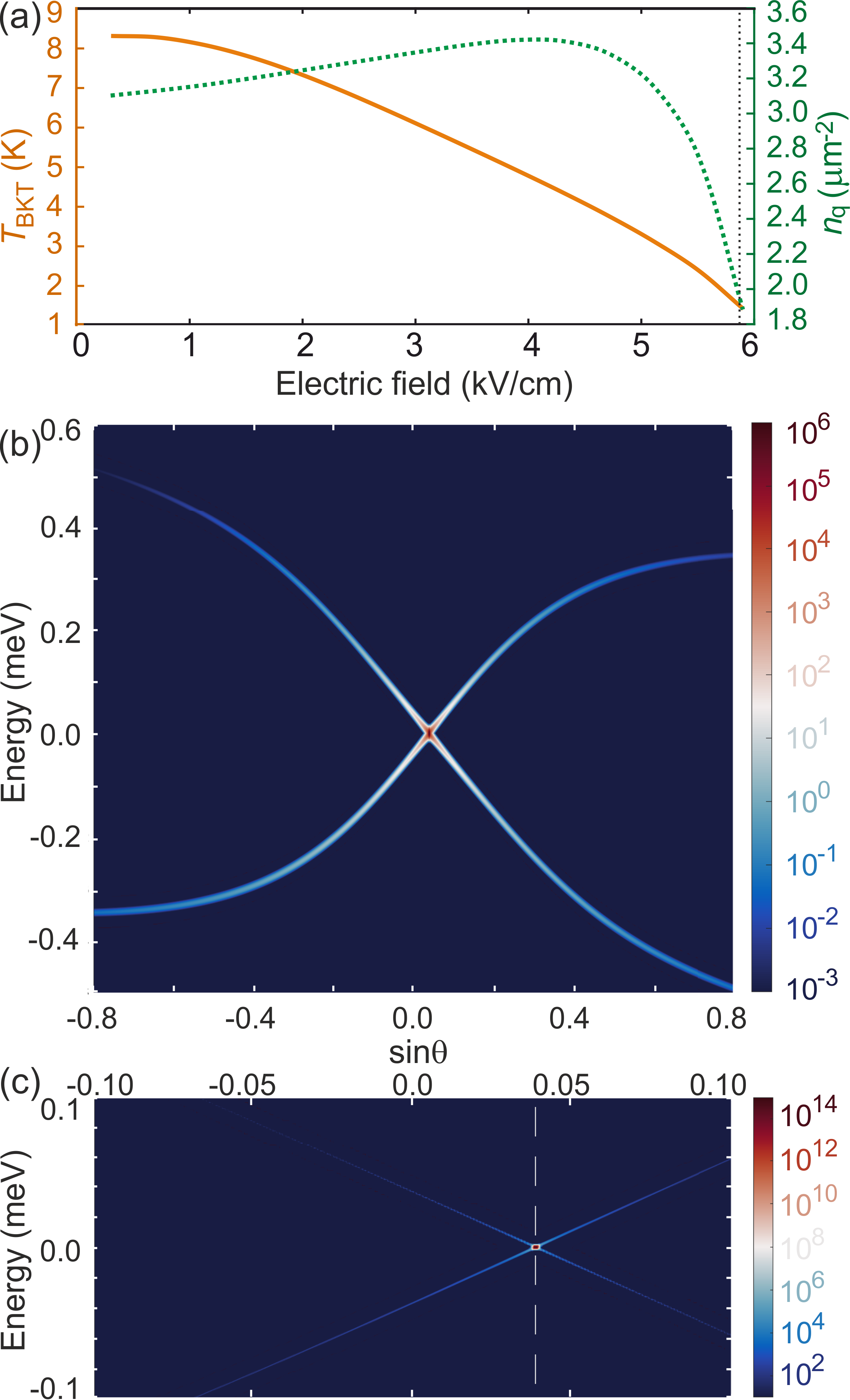}
\caption{(a) Critical temperature $T_{\rm BKT}$ (left axis, solid line) and the quasicondensate density $n_{\rm q}$ (right axis, dotted line) dependent on the electric field $E$, at $B=3$~T, $\hbar\Omega_0=6$~meV, $\Delta=10$~meV, $n=10^{10}$~cm$^{-2}$. (b--с) The spectral-angular dependence of the PL intensity according to Eq.~(\ref{ItpoS}) for $E=5.9$~kV/cm [corresponds to the vertical dotted line in (a)], $T=1$~K, with the energy origin taken at the level $\mu\approx E_G$. (b) PL from the Bogoliubov dispersion (without the condensate contribution): both the normal (thermally occupied) and the ghost (quantum-occupied) branches of the spectrum are visible; (c) with the condensate added, zoom-in on the region of the condensate momentum ${\bf K}_0$. The shift with respect to normal is $\theta_0=\text{arcsin}(0.04)\approx 2.3^\circ$ (marked by the vertical dashed line). The intensity colorscales in (b,c) are logarithmic, in arbitrary units, and normalised to the same quantity for both panels.
}
\label{fig5}
\end{figure}

Fixing the total density of polaritons to $n=10^{10}$~cm$^{-2}$, magnetic field $B=3$~T and the detuning $\Delta=10$~meV in the polariton-BEC regime, and considering the Bogoliubov excitations with the spectrum $\varepsilon_{\boldsymbol{k}}$ as noninteracting non-quasicondensate particles, we calculate the quasicondensate density $n_{\rm q} = n - \int n_{\boldsymbol{k}} d\boldsymbol{k}/(2\pi\hbar)^2$ varying the electric field strength $E$. The temperature at which $n_{\rm q}$ vanishes defines the critical temperature $T_{\rm BKT}$ of the BKT transition~\cite{prb104125301}. Fig.~\ref{fig5}{\bf a} shows both $n_{\rm q}$ and $T_{\rm BKT}$ against $E$. One notes that as long as the electric fields are weak enough to ensure that the polariton minimum of the dispersion is deeper than the exciton one, the critical temperature stays as high as a few K. However as soon as the growth of $E$ results in the competition of the two minima of the dispersion, the roton gap becomes small, leading to the quasicondensate density depletion and the quench of the critical temperature. The electric field $E=5.9$~kV/cm that is just below the transition to the exciton-BEC regime is marked in Fig.~\ref{fig5}{\bf a} by the vertical dotted line. In this borderline case, $T_{\rm BKT}=1.4$~K.

Finally, the zero-temperature anomalous Green's function of lower polaritons has the form
\begin{multline}\label{Fpo0}
F_{\boldsymbol{k}}(\omega)\equiv -i \!\int\limits_{-\infty}^{\infty} \!\!\!e^{i\omega t} \langle\hat{\rm T}[\hat a_{{\bf K}+\boldsymbol{k}}(t)\hat a_{{\bf K}-\boldsymbol{k}}(0)]\rangle dt = \\
-\frac{\hbar\,\mathcal{U}_{\boldsymbol{k}}n_0/n} {(\hbar\omega-\mathcal{A}_{\boldsymbol{k}})^2-(\mathcal{E}_{\boldsymbol{k}}- i\Gamma_{\boldsymbol{k}}/2)^2},
\end{multline}
with $\hat{\rm T}[...]$ denoting the chronological ordering and $\hat\alpha_{\boldsymbol{k}}(t)=\hat\alpha_{\boldsymbol{k}} e^{-i\varepsilon_{\boldsymbol{k}}t/\hbar}$ the annihilation operator of an excitation with the momentum $\boldsymbol{k}={\bf p}-{\bf K}$ in Heisenberg picture. The decay of excitations $\Gamma_{\boldsymbol{k}}\ge0$ which appears from the imaginary part of the anharmonic self-energy~\cite{AGD} is introduced in the denominator of Eq.~(\ref{Fpo0}) by hand, whereas the condensate fraction $n_0/n$ in the numerator appears from the more rigorous derivation in the formalism of unified Bogoliubov theory with quantum hydrodynamics.
The anomalous zero-temperature Green's function $F_{\boldsymbol{k}}(\omega)$ is not even with respect to both the momentum and frequency.

\section{Photoluminescence}
\label{B-opt}

In this section, we calculate the photoluminescence (PL) of Bose-condensed lower polaritons and the two-photon signal using the HBT scheme. According to the standard quantum-field diagrammatic formalism~\cite{tiop}, the intensity of spontaneous emission is defined as
\begin{equation}\label{Iq-tiop}
I=\frac VS\sum\limits_{\bf q}\!\int\limits_0^{\infty}\! dq_z\,\omega_{\vec{q}}\, \frac{|L_{\vec{q}}^{\lambda}|^2}{{\hbar}^2}(1-X_{\bf q}^2)
N\!\left(\!{\bf q},\omega_{\vec{q}}-\frac{\mu}{\hbar}\right)\!,
\end{equation}
where $\vec{q}=\{{\bf q},q_z\}$ is the 3D momentum of a photon leaving the cavity, $\omega_{\vec{q}}=c\sqrt{(|{\bf q}|^2+q_z^2)/\varepsilon}$ is its frequency, $L_{\vec{q}}^{\lambda}$ is the matrix element of dissipation: $$|L_{\vec{q}}^{\lambda}|^2=\frac{S\hbar^2c}{V{\tau}_{\vec{q}}^{\lambda} \sqrt{\varepsilon}},$$
with ${\tau}_{\vec{q}}^{\lambda}$ being the decay time of a photon towards the mode ($\vec{q}\lambda$), $\lambda$ is the condensate polarization, $S$ is the polariton system area, and $V\to\infty$ the volume of quantization. We use the condition $q_z>0$ to impose the impenetrability of the bottom mirror.

Since the renormalized exciton gap $E_G$ ($\approx1.5$~eV for GaAs) is large compared to the energies corresponding to polariton interaction and their coupling to light ($\sim$~meV), one can assume $\mu\pm\varepsilon_{\bf\mp p}\approx\mu\approx E_G$. Furthermore, considering the photon decay time independent of $\vec{q}$ and $\lambda$, namely, ${\tau}_{\vec{q}}^{\lambda}\approx\tau_0$, one gets after transformations for the intensity of the spectrally- and angle-resolved luminescence per unit area [see~(\ref{Iq-tiop})]:
\vspace{-2mm}
\begin{equation}\label{ItpoS}
\frac{I(\phi,\theta;\omega)}S=\frac{E_G}{\tau_0}
\frac{1-X_{\bf q}^2}{(2\pi\hbar/q_{\rm rad})^2}N\!\left({\bf q},\omega-\frac{\mu}{\hbar}\right).
\end{equation}
In (\ref{ItpoS}), the angular dependence enters via the lightcone boundary $q_x=q_{\rm rad}\sin\theta\cos\phi$, $q_y=q_{\rm rad}\sin\theta\sin\phi$.
Knowledge of the momentum-frequency particle distribution (\ref{npo}) allows us to calculate the spectral-angular distribution of the PL intensity~(\ref{ItpoS}). As only the polariton-BEC regime is accessible in luminescence, we address the situation at the verge of the transition to the exciton-BEC regime ($B=3$~T, $E=5.9$~kV/cm, $\Delta=10$~meV) corresponding to the excitation spectrum in Fig.~\ref{fig3b}{\bf a}. The PL distribution is plotted in Fig.~\ref{fig5}{\bf b} and {\bf c} against $\sin\theta$ (where $\theta$ is the emission angle along the $y$-axis, i.e. $q_x=0\Leftrightarrow\phi = \pi/2$, $-\pi/2$) for $T=1$~K which is just below $T_{\rm BKT}$ for these parameters, see Fig.~\ref{fig5}{\bf a}. The intensity distribution displays a clear asymmetry with respect to normal direction of emission. As the temperature is very low, one notes that the negative (ghost) branch of the Bogoliubov dispersion is occupied stronger compared to the normal (thermal) branch of excitations. Fig.~\ref{fig5}{\bf c} shows the magnified view of the low-momenta region. The shift of the condensate momentum $K_0$ from zero is clearly seen. We estimate the angle of condensate emission in this case to be $\theta_0\approx2.3^\circ$ in air.

\begin{figure}[t]
\includegraphics[width=\linewidth]{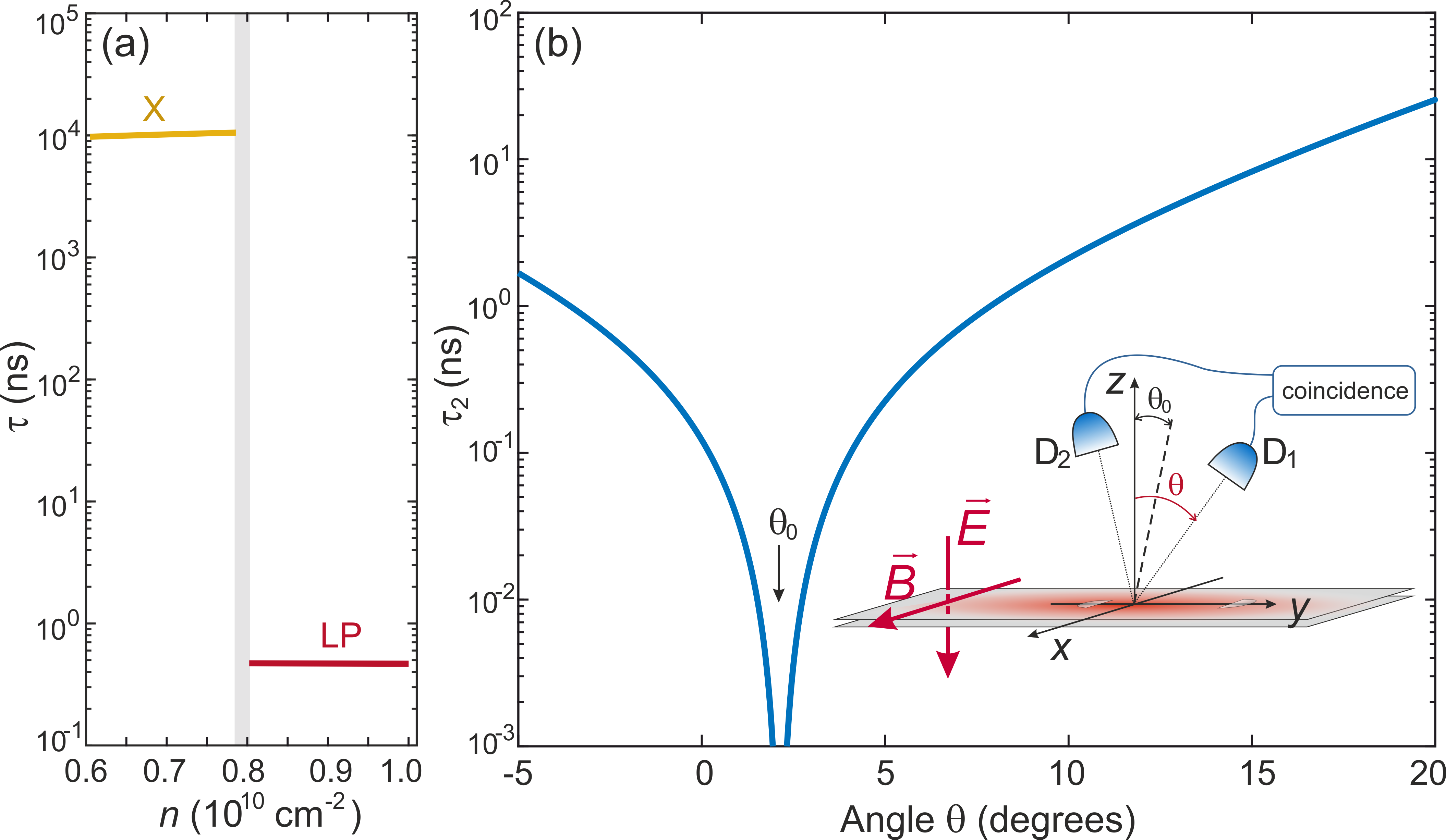}
\caption{(a) The system lifetime $\tau$ across the exciton-BEC --- polariton-BEC transition. The transition region is marked by the grey-shaded area. (b) Angular dependence of the two-photon decay time $\tau_2(\phi,\theta)$ of polaritons in the HBT coincidence experiment, the scheme of which is shown in the inset. Polar angles $\theta>\theta_0$ correspond to the detector (D$_1$) position at $\phi=\pi/2$, while $\theta<\theta_0$ corresponds to $\phi=-\pi/2$ (see detector D$_2$). The angle $\theta$ is counted along the direction of momentum ${\bf q}$ (here along $y$--axis), and the angle $\theta_0$ corresponds to the condensate momentum ${\bf K}_0$ (i.e. $\sin\theta_0 = K_0/q_{\rm rad}$).
Parameters are the same as for Fig.~\ref{fig5}, $\tau_0=10$~ps. For panel (b), $n=10^{10}$~cm$^{-2}$, $n_0/n=0.72$, $\tau=450$~ps.
}
\label{fig4}
\end{figure}

Integrating (\ref{ItpoS}) over the upper semisphere and over frequencies, we find the system lifetime~$\tau$:
\begin{equation}\label{tau}
\frac{1}{\tau}=\!\int\limits_0^{2\pi}\!d\phi\!\int\limits_0^{\pi/2}\!\sin\theta d\theta \! \int\limits_0^{\infty}\!d\omega\frac{I(\phi,\theta;\omega)}{nSE_G}.
\end{equation}
In Fig.~\ref{fig4}{\bf a}, we plot the lifetime (\ref{tau}) dependent on the total density $n$ across the transition, revealing the drastic drop of the radiative recombination rate at the polariton-BEC --- exciton-BEC transition (at the continuous decrease of $n$, $\tau$ changes from 473~ps to 11~$\mu$s). Such a quench of the decay happens due to the fact that in the exciton regime only the small part of the momentum space radiates, that which is responsible for the excitons coupling to light. It is striking that, in the polariton regime, the system lifetime of a low-quality cavity ($\tau_0=10$~ps) even at zero temperature is of the order of hundreds picoseconds, which justifies our assumption that the system is in thermal equilibrium.

Next, we consider the signal magnitude for the two-photon coincidences in the HBT experiment. The two-photon signal magnitude is defined as the number of photons counted by the first detector multiplied by the number of photons counted by the second one (in unit time per one polariton). Assuming that the two detectors are counting photons over the short time window $t_0$ which is still much longer than the excitation lifetime,
\begin{equation}\label{HBTdef}
\frac{1}{\tau_2}\!=\!\frac{\mathcal{N}(t_0)\!-\!\mathcal{N}(0)}{Nt_0},\quad
\mathcal{N}(t)\!\equiv\!\sum\limits_{\vec{q},\vec{q}^\prime}
\!\tr\!\rho_\mathcal{H}\hat N_{\vec{q}}(t)\hat N_{\vec{q}^\prime}(t).
\end{equation}
Here $\hat\rho_\mathcal{H}$ is the Heisenberg density matrix accounting for photon leak out of the cavity~\cite{tiop}, $\hat{N}_{\vec{q}}(t)=\hat{c}_{\vec{q}}^{\dagger}(t)
\hat{c}_{\vec{q}}(t)$ is the Heisenberg operator of number of photons in the mode $(\vec{q}\lambda)$, and the summation over $\vec{q}$, $\vec{q}^{\,\prime}$ is performed only over the photon frequencies and solid angle elements that correspond to the spatial orientation of the two detectors (see schematic illustration in the inset of Fig.~\ref{fig4}{\bf b}).

Making transformations in Eq.~(\ref{HBTdef}) and applying the Wick's theorem for the Heisenberg averages $\langle\dots\rangle_{\!\mathcal{H}}$ over the density matrix $\hat\rho_\mathcal{H}$, we obtain
\begin{multline}\label{<cccc>}
\langle\hat{c}_{\vec{q}}^{\dagger}(t)\hat{c}_{\vec{q}}(t)
\hat{c}_{\vec{q}^\prime}^{\dagger}(t)\hat{c}_{\vec{q}^\prime}(t) \rangle_{\!\mathcal{H}} = \langle\hat{c}_{\vec{q}}^{\dagger}(t) \hat{c}_{\vec{q}}(t)\rangle_{\!\mathcal{H}} \langle\hat{c}_{\vec{q}^\prime}^{\dagger}(t) \hat{c}_{\vec{q}^\prime}(t) \rangle_{\!\mathcal{H}} \\
+ \langle\hat{c}_{\vec{q}^\prime}^{\dagger}(t) \hat{c}_{\vec{q}}(t) \rangle_{\!\mathcal{H}} \langle\hat{c}_{\vec{q}}^{\dagger}(t) \hat{c}_{\vec{q}^\prime}(t)\rangle_{\!\mathcal{H}} + \langle\hat{c}_{\vec{q}}^{\dagger}(t)\hat{c}_{\vec{q}}(t) \hat{c}_{\vec{q}^\prime}^{\dagger}(t) \hat{c}_{\vec{q}^\prime}(t) \rangle_{\!\mathcal{H}}^{\!\rm c} \\
+ \langle\hat{c}_{\vec{q}^\prime}^{\dagger}(t)\hat{c}_{\vec{q}}^{\dagger}(t) \rangle_{\!\mathcal{H}}\langle\hat{c}_{\vec{q}}(t)\hat{c}_{\vec{q}^\prime}(t) \rangle_{\!\mathcal{H}}.
\end{multline}
Eq.~(\ref{<cccc>}) contains four terms. The first term, which is quadratic with respect to luminescence, does not have any angular distribution and is proportional to $t^2$. In the two-photon coincidence scheme it can be omitted. The second term is proportional to $t$ and possesses an angular directionality ($\propto\delta_{\bf qq^\prime}$). However, in the case the spatial orientation of the detector does not correspond to ${\bf q=q}^\prime$, this term will be absent in the signal~(\ref{HBTdef}). The third term represents the connected four-photon vertex which, even while being $\sim t$, does not have any angular directionality (i.e. it only adds noise to the signal). Finally, the fourth term is also $\sim t$ and possesses the angular directionality of the form $\propto\delta_{{\bf q'},2{\bf K-q}}$. Therefore, if the orientation of the detectors is tuned to this term, it will be the only one contributing to the two-photon signal~(\ref{HBTdef}).

Accounting for the last term in Eq.~(\ref{<cccc>}), we obtain the HBT signal magnitude in unit solid angle
\begin{equation}\label{HBTfin}
\frac1{\tau_2(\phi,\theta)}=\frac{(1-X_{\bf q}^2)(1-X_{2{\bf K-q}}^2)}
{(2\pi\hbar\tau_0/q_{\rm rad})^2n\cos\theta}I_2,
\end{equation}
with
\begin{equation}
I_2 = \!\!\int\limits_{-\infty}^{\infty}\!\!\frac{d\omega}{2\pi}
\left|F_{\bf q-K}(\omega)\right|^2 = \frac{\mathcal U^2({\bf q-K})n_0^2\tau_{\bf q-K}}{2\mathcal{E}_{\bf q-K}^2+\hbar^2/2\tau_{\bf q-K}^2}
\end{equation}
and $\tau_{\boldsymbol{k}}=\hbar/\Gamma_{\boldsymbol{k}}$ denoting the excitations lifetime. In the polariton-BEC regime, at small momenta (in the vicinity of the condensate) it is defined predominantly by the system lifetime~\cite{prl099140402}: $\tau_{\boldsymbol{k}}\approx\tau$. In the exciton regime, $\tau$ can be very long (as shown in Fig.~\ref{fig4}{\bf a}), and for realistic parameters the radiation channel is not dominant.

In Fig.~\ref{fig4}{\bf b}, we show the inverse HBT signal magnitude, plotting the angular dependence of the two-photon decay time $\tau_2(\phi,\theta)$  [according to (\ref{HBTfin})] at small angles $\theta$ (in the vicinity of the condensate) in the polariton-BEC regime, i.e. when the dominant decay channel is luminescence. It is clearly seen that the signal grows drastically at small $\theta$, with $\tau_2(\phi,\theta)$ reaching sub-nanosecond scales and less. When changing the angle, the signal decreases as $\tau_2(\phi,\theta)$ grows, while still having the order of nanoseconds. It is noteworthy that not only the angular change of $\tau_2$ is an observable effect but also that the signal~(\ref{HBTfin}) occurs only in the case when the anomalous Green's function $F_{\boldsymbol{k}}(\omega)$ is nonzero. Since this happens only when the system features a Bose condensate, the measurement of the signal in the HBT scheme can be used as a direct evidence of the existence of the dipolariton BEC.

\section{Сonclusions} \label{Concl}

We propose a realization for (quasi-)equilibrium long-living BEC of dipolaritons in a wide single quantum well in an optical microcavity.
By combining the in-plane magnetic and transverse electric fields, we demonstrate the field-controlled appearance of the two energy-competing minima in the particle dispersion, in contrast to both the usual $p^2/2m$ paraboloid and a more sophisticated non-parabolic spectrum of lower polaritons in the absence of external fields. The energy competition of these two minima---polaritonic and excitonic---manifests in the appearance of an abrupt transition from the polariton BEC to the exciton BEC (and vice versa) upon a continuous change of one parameter: either the total density or photon-exciton energy detuning, or, alternatively, the electric field strength. We show that this transition displays the signature of a first-order phase transition, with the pressure being continuous while its derivative with respect to chemical potential experiencing a jump. Under these conditions, the optically-dark exciton mode with microsecond decay times becomes achievable.

Furthermore, we developed the many-body theory of dipolaritons in crossed fields accounting for the combined effect of the new peculiar dispersion and the extended-range dipole-dipole interactions. After having obtained the dressed effective Hamiltonian of the system, we performed the Bogoliubov diagonalization that reveals two substantially different, anisotropic excitation spectra in the two condensation regimes, both of them displaying non-symmetric roton-maxon softening in momentum regions away from the condensate. We note that in both regimes, the Bose condensation occurs at a non-zero in-plane momentum ${\bf K_0}$, and that the (anisotropic) Landau critical velocity in all in-plane directions is much smaller than the sound velocity defined at ${\bf p}\to{\bf K_0}$.

Our theory which stitches the Bogoliubov approach with that of quantum hydrodynamics accounts for both the lack of parity and of Galilean invariance in the system, and provides estimates for all the relevant parameters, such as the condensate fraction, momentum-frequency distribution, radiative lifetime, the BKT transition temperature, and the anomalous Green's function.
The spectral-angular distribution of the PL intensity indicates that in the regime of polariton (radiative) BEC, the main luminescence peak is deviated from normal direction by a detectable angle $\theta_0$, and both the normal and ghost branches of the dispersion of elementary excitations are anisotropic. The calculated dependence of the two-photon decay time using the Hanbury Brown--Twiss coincidence scheme, in the case when the two detectors are placed symmetrically with respect to $\theta_0$ along the direction perpendicular to the applied magnetic field, shows a sharp angular dependence of the HBT signal magnitude.

We hope that this work will stimulate experimental realisations of dipolariton BECs, including in cross fields and under the conditions of suppressed radiative decay even in low-finesse microcavities. The controllable transition between the bright and dark BECs can be used to control photoluminescence and light-matter transport. The presence of a second (roton-like) minimum already in the single-particle polariton dispersion paves the way to on-demand realisation of such long-sought roton-maxon phenomena in the excitation spectra as the density waves, crystallization and supersolids.

\section*{Acknowledgments}
The work on the elementary excitations (Sec.~\ref{ham}--\ref{B-approx}) is being developed within the Russian Science Foundation (RSF) Project BEL~23--42--10010. Quantum-hydrodynamic theory (Sec.~\ref{sec_corr}--\ref{B-opt}) is partially funded by the Russian Foundation for Basic Research (RFBR) within the Project No.~21--52--12038. N.~S.~V. acknowledges the financial support of the NRNU MEPhI Priority 2030 program.\\

\appendix
\section{The bare polariton Hamiltonian in the strong-coupling regime}
\label{AppH}

In this Appendix, we derive the Hamiltonian (\ref{Hexph}) starting from the electron-hole Hamiltonian in presence of static in-plane magnetic and transverse electric fields.
We assume that the in-plane magnetic field $B{\bf e}_x$ is described by the vector potential ${\bf A}_B(z)=-Bz{\bf e}_y$, while the out-of-plane electric field is given by $-E\vec{\mathrm{e}}_z$. In the effective mass approximation,
\begin{widetext}
\begin{multline}\label{HehA}
\hat{\mathcal{H}} =\! \int\!\! d\vec{r}_e\hat{\psi}_e^\dag(\vec{r}_e)
\!\left[E_g + \frac{1}{2m_e}\!\left(-i\hbar\vec\nabla_e + \frac{e}{c} {\bf A}_B(z_e) + \frac{e}{c}\hat{\vec{\textrm{A}}}(\vec{r}_e)\right)^{\!\!2} + W_e(z_e)- eEz_e\right]\!\hat{\psi}_e(\vec{r}_e) \\
+ \!\int\!\! d\vec{r}_h\hat{\psi}_h^{\dagger}(\vec{r}_h)\!\left[\frac{1}{2m_h}\! \left(-i\hbar\vec\nabla_h - \frac{e}{c}{\bf A}_B(z_h) - \frac{e}{c}\hat{\vec{\textrm{A}}}(\vec{r}_h)\right)^{\!\!2} + W_h(z_h) + eEz_h\right]\!\hat{\psi}_e(\vec{r}_h) + \hat{H}_{\rm eh}^\prime \\
+  \frac{e^2}{2\varepsilon} \sum\limits_{i,j=e,h} \int \!\! d\vec{r}_{\!i} d\vec{r}_{\!j} \frac{\hat{\psi}_i^\dag(\vec{r}_{\! i})\hat{\psi}_j^\dag(\vec{r}_{\!j})
\hat{\psi}_j(\vec{r}_{\!j})\hat{\psi}_i(\vec{r}_{\!i})}{\left| \vec{r}_{\!i}-\vec{r}_{\!j}\right|}
+ \hat{H}_{\rm ph} + \!\int \!\!d\vec{r} \left(\hat{\psi}_e(\vec{r})\frac{iE_g}{\hbar c}\, \vec{\mathrm{d}}_{\mathrm{vc}}\,\hat{\vec{\mathrm{A}}}(\vec{r})\hat{\psi}_h(\vec{r}) + \text{h.c.}\right),
\end{multline}
\end{widetext}
where $\hat{\psi}_{e(h)}(\vec{r})$ is the Fermi field operator of an electron (hole) with the spin projection that participates in the polariton BEC.
The interband dipole moment $\vec{\rm d}_{\rm vc} =\int u_{\rm v}^*(\vec{r})\:e\vec{r}\:u_{\rm c}(\vec{r})d\vec{r}$ is defined by the Bloch functions of the valence $u_{\rm v}(\vec{r})$ and conduction $u_{\rm c}(\vec{r})$ bands. The field operator of photons
\begin{equation}\label{Ar}
\hat{\vec{\mathrm{A}}}(\vec{r}) \!=\!\! \sum\limits_{\bf p}
\!\sqrt{\frac{2\pi\hbar^2c^2}{\varepsilon\hbar\omega_{\bf p}S}}\!\left[
\hat c_{\bf p}e^{i{\bf p \cdot r}/\hbar}\varphi(z)\vec{\mathrm{e}}_{\bf p} \!+\! \text{h.c.}\right] \!+ \hat{\vec{\mathrm{A}}}^\prime(\vec{r})
\end{equation}
is taken in the gauge $\diverg\hat{\vec{\mathrm{A}}}(\vec{r})=0$ and is defined via the 2D photon annihilation operator $\hat{c}_{\bf p}$ and the transverse-quantized wavefunction $\varphi(z)$ normalized according to $\int_{-\infty}^{\infty}|\varphi(z)|^2dz=1$. $\vec{\rm e}_{\bf p}$ is the polarization vector corresponding to the mode which features the polariton BEC.

The Hamiltonian of free electromagnetic field (in the cavity) in the third line of Eq.~(\ref{HehA}) is
\begin{equation}\label{Tph}
\hat H_{\rm ph} = \sum\limits_{\bf p}
\hbar\omega_{\bf p}\hat c_{\bf p}^{\dagger}\hat c_{\bf p} + \hat{H}_{\rm ph}^\prime.
\end{equation}
Both (\ref{Ar}) and (\ref{Tph}) contain the summation over discrete 2D momenta ${\bf p}=(2\pi\hbar/L){\bf j}$, where $L=\sqrt S$ and ${\bf j}\in\mathbb Z^2$ is a 2D integer-valued vector. The terms $\hat{\vec{\mathrm{A}}}^\prime(\vec{r})$, $\hat{H}^\prime_{\rm ph}$ and $\hat{H}_{\rm eh}^\prime$ contain the photon modes and polarizations, or, respectively, the electron (hole) fields and spin projections that are not participating in the polariton BEC and present little interest.

The electron-hole Hamiltonian (\ref{HehA}), when transiting to the exciton picture, can be simplified using the following considerations.
As Bose condensation occurs only on one (spontaneously chosen) polarization branch, the occupation of the photon mode with the opposite polarization is small, as well as the occupation of non-condensate exciton spin branches. The interaction of excitons with other, non-condensate, cavity modes is negligible~\cite{footnote1}. At the same time, interaction of the condensate particles with all the other incoherent excitons present in the system provides in the leading order of perturbation theory only the blueshift of the exciton chemical potential~\cite{prb104125301} (i.e. the renormalization of the semiconductor gap $E_g$), and does not contribute to the particle pair interaction~\cite{prb099085108}. Due to these reasons, the terms $\hat{\vec{\mathrm{A}}}^\prime(\vec{r})$, $\hat{H}^\prime_{\rm ph}$ and  $\hat{H}_{\rm eh}^\prime$ in Eqs. (\ref{HehA})--(\ref{Tph}) can be safely omitted.
Furthermore, the intraband interaction of charge carriers with photons leads to negligible virtual jumps of an electron (hole) up or down within the corresponding band, hence in the first two lines of (\ref{HehA}) containing the intraband single-particle operators one can set $\hat{\vec{\mathrm{A}}}(\vec{r})=0$.

Since the characteristic energies of the exciton system, such as the temperature and chemical potential, are small compared to the energy needed to excite internal exciton degrees of freedom, in the exciton particle operator we account only for the centre-of-mass motion. Neglecting also the composite-boson nature of excitons~\cite{combescot} due to the assumed regime of strong coupling, we can follow the standard second quantization procedure and truncate the full Hilbert space of states of the electron-hole-photon system, so as to consider the subspace corresponding only to the ground state of transverse quantization and $1s$--state of the relative electron-hole motion, as well as to only the condensate cavity photon mode, exciton spin branch, and photon polarization.

As a result, the Hamiltonian (\ref{HehA}) after some algebra acquires the form of Eq.~(\ref{Hexph}), with the exciton annihilation operator defined as
\begin{equation}\label{Qp}
\hat{Q}_{\bf p}=\int d\vec{r}_ed\vec{r}_h \phi_{\bf p}(\vec{r}_e,\vec{r}_h) \hat{\psi}_e(\vec{r}_e)\hat{\psi}_h(\vec{r}_h),
\end{equation}
and the electron--hole Coulomb interaction $U_0(\vec{r}_e,\vec{r}_h,\vec{s}_e,\vec{s}_h)$ in (\ref{U0pkk'}) given by
\begin{equation}\label{U0rseh}
U_0 
\!=\! \frac{e^2}{\varepsilon} \!\! \left(\! \frac{1}{|\vec{r}_e \!-\! \vec{s}_e|} \!+\! \frac{1}{|\vec{r}_h \!-\! \vec{s}_h|} \!-\! \frac{1}{|\vec{r}_e \!-\! \vec{s}_h|} \!-\! \frac{1}{|\vec{r}_h \!-\! \vec{s}_e|}\!\right)\!.
\end{equation}

\section{Electron and hole wavefunctions in crossed fields}\label{AppMin}

The Hamiltonian~(\ref{Hrerh}) of the exciton eigenvalue problem, rewritten in terms of the in-plane and transverse coordinates, has the form:
\begin{multline}\label{H0_cm}
\hat{H}_0 = E_g - \frac{\hbar^2}{2m_{\rm ex}}\nabla_{\!\bf r}^2 - \frac{\hbar^2}{2\mu_{eh}}\nabla_{\!\!\boldsymbol{\varrho}}^2 - \frac{\hbar^2}{2m_e}\frac{\partial^2}{\partial z_e^2} - \frac{\hbar^2}{2m_h}\frac{\partial^2}{\partial z_h^2} \\+ i\hbar\frac{eB}{cM}(z_e-z_h)\frac{\partial}{\partial r_y} + i\hbar\frac{eB}{c\mu_{eh}}\frac{m_hz_e+m_ez_h}{m_{\rm ex}} \frac{\partial}{\partial\varrho_y} \\+ \frac{e^2B^2}{2m_ec^2}z_e^2 + \frac{e^2B^2}{2m_hc^2}z_h^2 + W_e(z_e) + W_h(z_h)\\ - eE(z_e-z_h) - \frac{e^2}{\varepsilon\sqrt{\varrho^2+(z_e-z_h)^2}},
\end{multline}
with $\mu_{eh} = m_em_h/m_{\rm ex}$ denoting the electron-hole reduced mass. Substituting the ansatz (\ref{phi_trial}) in $\hat H_0\phi_{\bf p}= E_{\bf p}\phi_{\bf p}$ yields the Eq.~(\ref{Schred}) for the electron and hole wavefunctions in the wide QW in presence of external fields.

Solution of this equation and the minimization problem~(\ref{<phi|H0|phi>})--(\ref{phi_trial}) can be simplified upon consideration of the physical parameters. For GaAs-based microcavities, the typical particle (polariton) densities are of the order $\sim10^{10}$~cm$^{-2}$. Then, for $p_y\sim2\pi\hbar\sqrt n$ and $B\sim2$~T, one estimates the addition $Bp_y/m_{\rm ex}c$ to the electric field $E$ in (\ref{Schred}) to be of the order of~$0.1$~kV/cm. Even for the weakest electric field that we consider here ($E\sim 5$~kV/cm) it does not affect the $z$--profile of the functions $f_{e,h}(z,p_y,k_y)$~\cite{jetpl0830553}. Therefore when calculating the exciton dipole moment $d$ or similar integrals, we may set $p_y\approx0$
and $f_{e,h}(z,p_y,k_y)\approx f_{e,h}(z,0,k_y)$. Then Eqs.~(\ref{Schred}) yield immediately
\begin{align}
\mathscr{E}_e(p_y,k_y) + & \mathscr{E}_h(p_y,k_y) =  \nonumber \\
& \mathscr{E}_e(0,k_y) + \mathscr{E}_h(0,k_y)-\frac{Bp_y}{m_{\rm ex}c}d(0,k_y), \label{Ez(py)}
\end{align}
where $d(p_y,k_y)\equiv e\int\limits_{-\infty}^\infty(z_e-z_h) |\phi_{\bf p}(\vec{r}_e,\vec{r}_h)|^2d\vec{r}_ed\vec{r}_h$.

The second consideration is that for wide GaAs QWs, the realistic electron-hole separation and mass ratio can be estimated as $\bar z_e-\bar z_h\sim12$~nm and $m_e/m_h\approx1/6$, where $\bar z_{e(h)}=\int z_{e(h)}|\phi_{\bf p}(\vec{r}_e ,\vec{r}_h)|^2d\vec{r}_ed\vec{r}_h$ is the average electron (hole) coordinate in the growth direction. At the same time, from numerics one sees that for a single QW in electric field $\bar z_e\sim\sqrt{m_e/m_h}(-\bar z_h)$ is a good estimate (we assume that the coordinate $z=0$ corresponds to the center of the QW, as illustrated in Fig.~\ref{fig1}{\bf a}). Hence the addition $(eB/c)(\bar z_e/m_e+\bar z_h/m_h)k_y$ to the r.~h.~s. of (\ref{Ez(py)}) can be approximately estimated as 0.03~meV, while the $z$--profile of the functions $f_{e,h}(z,0,k_y)$ is defined by a much larger value $eE(\bar z_e-\bar z_h)\sim10$~meV. It is therefore justified to restrict our consideration only to the leading order in the ratio of these two quantities, and take $f_{e,h}(z,0,k_y)\approx f_{e,h}(z,0,0)$ when calculating the mean values $\bar z_e$ and $\bar z_h$. Then
\begin{align}
\mathscr{E}_e(0,k_y) + \mathscr{E}_h(0,k_y) &- \frac{Bp_y}{m_{\rm ex}c}d(0,k_y) = \nonumber \\
& \mathscr{E}_e + \mathscr{E}_h - \frac{q_0k_y}{\mu_{eh}}-\frac{Bp_yd}{m_{\rm ex}c}, \label{EeEh(ky)}
\end{align}
where
\begin{equation}\label{q0}
q_0 = \frac{eB}c\int\limits_{-\infty}^{\infty}\frac{m_hz_e+m_ez_h}{m_{\rm ex}} |f_e(z_e)f_h(z_h)|^2dz_edz_h
\end{equation}
and $\mathscr{E}_e=\mathscr{E}_e(0,0)$, $\mathscr{E}_h=\mathscr{E}_h(0,0)$, $f_e(z)=f_e(z,0,0)$, $f_h(z)=f_h(z,0,0)$.

\section{Dressing of the exciton-exciton interaction}\label{AppU}

Using the wavefunction (\ref{phip}), we can rewrite the last term in Eq.~(\ref{Hexph}) describing the bare exciton-exciton interaction via the exciton Bose field operators
$$\hat{Q}({\bf r}) = \frac{1}{\sqrt{S}} \sum\limits_{\bf p}e^{i{\bf p\cdot r}/\hbar} \hat{Q}_{\bf p}$$
and the pair interaction potential [see (\ref{U0rseh})]
\begin{align}
U_0({\bf r} \!-\! {\bf r}^\prime) \!=\! \!\int\!\! d\boldsymbol{\varrho} d\boldsymbol{\varrho}^\prime &dz_e dz_h dz_e^\prime dz_h^\prime U_0(\vec{r}_e,\vec{r}_h,\vec{r}_e^{\,\prime},\vec{r}_h^{\,\prime})\times \nonumber \\
& |\phi(\boldsymbol{\varrho},z_e,z_h) \phi(\boldsymbol{\varrho}^\prime,z_e^\prime,z_h^\prime)|^2, \label{U0rho}
\end{align}
where
\begin{equation}\label{phi_internal}
\phi(\boldsymbol{\varrho},z_e,z_h) \!=\! \frac{2\lambda_0}{\sqrt{2\pi}} e^{iq_0\varrho_y/\hbar}\!\exp(-\lambda_0\sqrt{x^2 \!+\! y^2})f_e(z_e)f_h(z_h)
\end{equation}
is the wave function of internal exciton degrees of freedom [i.e. $\phi_{\bf p}(\vec{r}_e,\vec{r}_h) = (1/\sqrt{S})\exp\{i{\bf p}\!\cdot\!{\bf r}\}\phi(\boldsymbol{\varrho},z_e,z_h)$], which is normalized according to $\int|\phi(\boldsymbol{\varrho},z_e,z_h)|^2d\boldsymbol{\varrho}dz_edz_h=1$. Namely, in the interaction term of the Hamiltonian~(\ref{Hexph}),
\begin{multline}\label{U0}
\frac{1}{2}\sum\limits_{\bf p,q,q^\prime} U_0({\bf p,q,q^\prime})
\hat{Q}_{\bf q}^\dag\hat{Q}_{\bf q^\prime}^\dag\hat{Q}_{\bf q^\prime+p}\hat{Q}_{\bf q-p} = \\
\frac{1}{2}\int U_0({\bf r}-{\bf r}^\prime) \hat{Q}^\dag({\bf r})\hat{Q}^\dag({\bf r}^\prime)\hat{Q}({\bf r}^\prime) \hat{Q}({\bf r})d{\bf r}d{\bf r}^\prime,
\end{multline}
we separate the short-range part of the dipole-dipole interaction (including the singularity) as
\begin{align}
\hat U_{\rm ex} & 
= \!\int\! \epsilon_0[\hat Q^{\dagger}({\bf r})\hat Q({\bf r})]d{\bf r} \label{U1B} \\ 
+ & \frac{1}{2}\!\int[U_0({\bf r\!-\!s}) \!-\! g_0\delta({\bf r \!-\! s})] \hat{Q}^\dag({\bf r})\hat{Q}^\dag({\bf s})\hat{Q}({\bf s})\hat{Q}({\bf r})d{\bf r}d{\bf s}. \nonumber 
\end{align}
In (\ref{U1B}), 
the first term (the short-range part) is taken in the local density approximation and accounts for many-body effects~\cite{prb087205302}, with $\epsilon_0(n_{\rm ex})$ being the part of the free energy per unit area (in a uniform system) responsible for exciton-exciton interaction, $n_{\rm ex}=\int\langle\hat{Q}^\dag({\bf r})\hat{Q}({\bf r})\rangle d{\bf r}/S$, and the averaging $\langle...\rangle$ is taken over the equilibrium density matrix of the polariton system~\cite{e0NormalOrdering}. The second term, on the other hand, corresponds to the first Born approximation for the quantity $U_0({\bf r})-g_0\delta({\bf r})$ which is considered to be small enough (here $g_0=\int U_0({\bf r})d{\bf r}$ is the bare interaction constant), and describes the long-range and (or) extended-range effects of the bare pairwise potentials. The subtraction of the $\delta$--contribution is performed in order to achieve the integrability of the inter-exciton potential at ${\bf r}=0$ in the strict dipolar limit~\cite{prb095245430}.

\begin{figure}[b!]
\includegraphics[width=\linewidth]{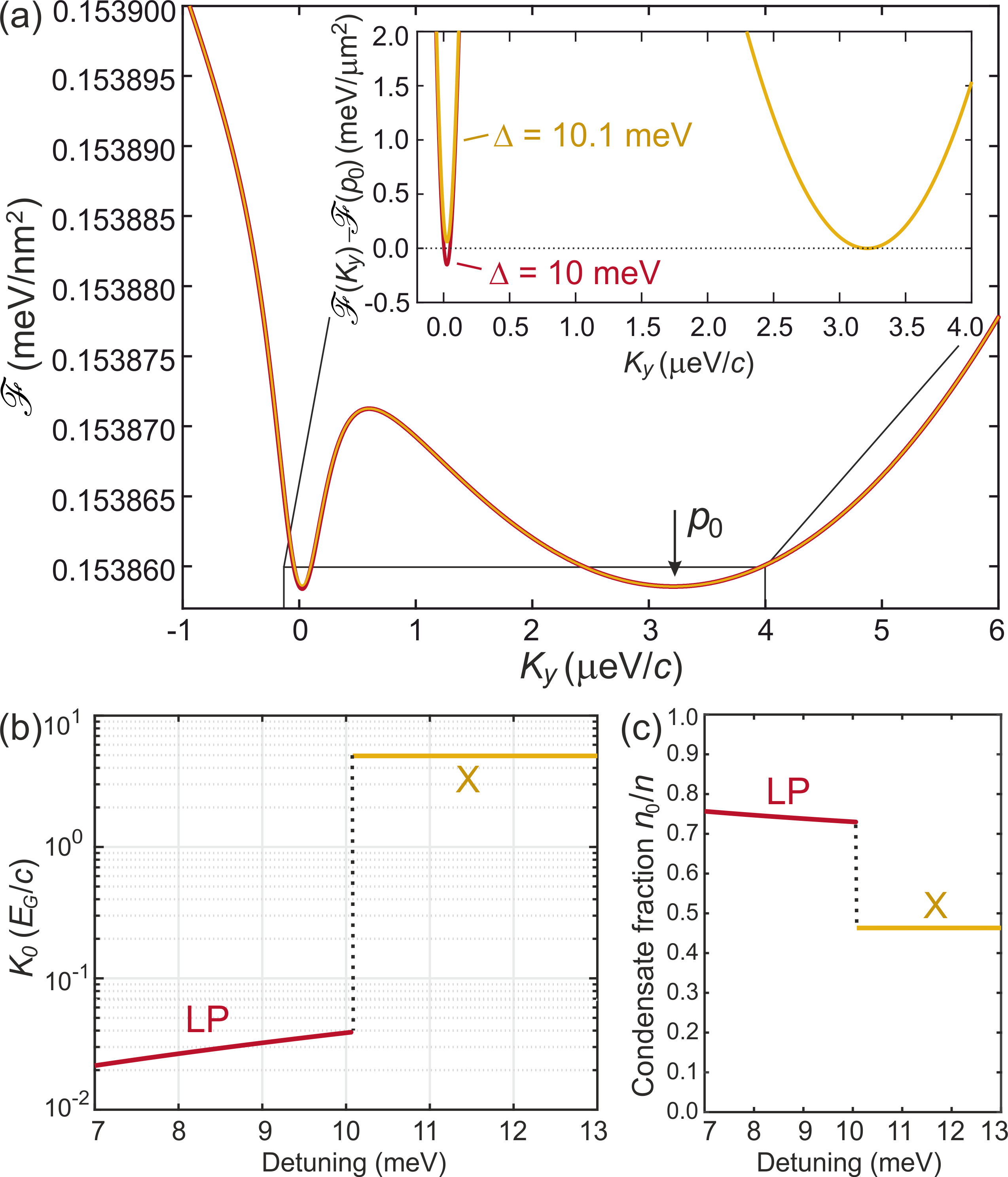}
\caption{Transition from the lower-polariton to the exciton BEC while changing the detuning. (a) The free energy functional according to Eq.~(25) of the main text at $\Delta = 10.0$~meV (the red line) and $\Delta = 10.1$~meV (the yellow line). The inset shows a magnified view of the two minima, indicating the `polariton' minimum becoming shallower than the `excitonic' one at $p_0$ when changing the detuning. (b) Condensate momentum ${\bf K} = {\bf K}_0$ (absolute value) across the transition. At $\Delta = 10.0$~meV, in the polariton-BEC regime $K_0 = 59.21~\text{meV}/c$, at $\Delta = 10.1$~meV, in the exciton-BEC regime $K_0 = 7595.1~\text{meV}/c$, where $c$ is the speed of light in vacuum. (c) Condensate fraction across the transition. For all panels, $B=3$~T, $E=5.9$~kV/cm, $d/e = 8.5$~nm, total density $n = 10^{10}$~cm$^{-2}$, $\hbar\Omega_0=6$~meV at $E=B=0$ and 3.5~meV in the applied $E$ and $B$. The renormalized exciton gap $E_G = 1538.46$~meV.  The dotted black lines indicate the transition.
}
\label{fig_S1}
\end{figure}

For simplicity, we find the free energy per unit area $\epsilon_0(n_{\rm ex})$ from the {\it ab initio} simulations of the strict 2D dipoles at $T=0$ without coupling to light~\cite{ssc144000399}:
\begin{equation}\label{e0nex}
\epsilon_0(n_{\rm ex}) \!=\! \frac{d^2}{\varepsilon r_D^5} a_1e^{(1+a_2) \ln\!u + a_3\ln^2\!u + a_4\ln^3\!u + a_5\ln^4\!u},
\end{equation}
where $u=n_{\rm ex}r_D^2$ is the dimensionless density, $r_D=m_{\rm ex}d^2/\hbar^2\varepsilon$, and the coefficients $a_1=9.218$, $a_2=1.35999$, $a_3=0.011225$, $a_4=-0.00036$, and
$a_5=-0.0000281$ correspond to the fitting in the interval $1/256\le u\le8$.

\section{Polariton-BEC --- exciton-BEC transition upon changing the photon-exciton detuning}\label{AppDet}

Here, we provide the results of investigation of the transition between the two BEC regimes at a fixed total density $n = 10^{10}$~cm$^{-2}$ while changing the photon-exciton detuning $\Delta$. In Fig.~\ref{fig_S1}{\bf a}, we plot an exemplary dependence of the free energy given by Eq.~(25) in the main text, on magnetic momentum ${\bf K}$.
The minimization procedure allows to find the condensate momentum ${\bf K}_0$ for each $E$ and $B$ depending on the total density $n$ and detuning $\Delta$. In Fig.~\ref{fig_S1}{\bf b}, we plot an example of such a dependence for $B=3$~T, $E=5.9$~kV/cm (at $\hbar\Omega_0=6$~meV). In both Fig.~\ref{fig_S1}{\bf a} and {\bf b}, a clear transition is seen at $\Delta\approx10.07$~meV . With the growth of $\Delta$, the resting (${\bf K=K}_0$) superfluid Bose-condensed system of excitons with a given total density (here $n=10^{10}$~cm$^{-2}$), due to the energy considerations---according to Eq.~(24) of the main text---undergoes a transition from its polariton minimum of the free energy to the exciton one (shown in the inset of Fig.~\ref{fig_S1}{\bf a}). Fig~\ref{fig_S1}{\bf c} shows the drop of the condensate fraction $n_0/n$ across the exciton-BEC --- polariton-BEC transition at a fixed total density $n$.

We note that since changing the electric field strength $E$ would alter the Rabi splitting $\hbar\Omega$ given by Eq.~(2) in the main text, it would result in effective change of the detuning with respect to $\hbar\Omega$ and hence of the exciton fraction in dipolaritons. Thus we conclude that one could observe a transition similar to described in Fig.~\ref{fig_S1} upon changing the electric field, while keeping the detuning $\Delta$ fixed.



\begin{thebibliography}{99}
\bibitem{you}
L. You and M. Marinescu, ``Prospects for $p$-wave paired Bardeen-Cooper-Schrieffer states of fermionic atoms'', Phys. Rev. A {\bf 60}, 2324 (1999).

\bibitem{shlyap1}
M. A. Baranov, M. S. Mar'enko, Val. S. Rychkov, and G. V. Shlyapnikov, ``Superfluid pairing in a polarized dipolar Fermi gas'', Phys. Rev. A {\bf 66}, 013606 (2002).

\bibitem{kuzirski}
D. H. J. O'Dell, S. Giovanazzi, and G. Kurizki, ``Rotons in Gaseous Bose-Einstein Condensates Irradiated by a Laser'', Phys. Rev. Lett. {\bf 90}, 110402 (2003).

\bibitem{shlyap2}
L. Santos, G. V. Shlyapnikov, and M. Lewenstein, ``Roton-Maxon Spectrum and Stability of Trapped Dipolar Bose-Einstein Condensates'', Phys. Rev. Lett. {\bf 90}, 250403 (2003).

\bibitem{lewenstein}
K. G'{o}ral, L. Santos, and M. Lewenstein, ``Quantum Phases of Dipolar Bosons in Optical Lattices'', Phys. Rev. Lett. {\bf 88}, 170406 (2002).

\bibitem{bloch}
M. Greiner, O. Mandel, T. Esslinger, T. W. H\"{a}nsch, and I. Bloch, ``Quantum phase transition from a superfluid to a Mott insulator in a gas of ultracold atoms'', {\it Nature} (London) {\bf 415}, 39--44 (2002).

\bibitem{giovanazzi}
S. Giovanazzi, D. O'Dell, and G. Kurizki, ``Density Modulations of Bose-Einstein Condensates via Laser-Induced Interactions'', Phys. Rev. Lett. {\bf 88}, 130402 (2002).

\bibitem{pfau}
M. Guo, F. B\"{o}ttcher, J. Hertkorn, J.-N. Schmidt, M. Wenzel, H. Peter B\"{u}chler, T. Langen and T. Pfau, ``The low-energy Goldstone mode in a trapped dipolar supersolid'', {\it Nature} (London) {\bf 574}, 386 (2019).

\bibitem{pfau_rev}
F. B\"{o}ttcher , J.-N. Schmidt, J. Hertkorn, K. S. H. Ng,
S. D. Graham, M. Guo, T. Langen and T. Pfau, ``New states of matter with fine-tuned interactions: quantum droplets and dipolar supersolids'', Rep. Prog. Phys. {\bf 84}, 012403 (2021).

\bibitem{ketterle}
S. Inouye, M. R. Andrews, J. Stenger, H.-J. Miesner, D. M. Stamper-Kurn and W. Ketterle, ``Observation of Feshbach resonances in a Bose–Einstein condensate'', {\it Nature} (London) {\bf 392}, 151--154 (1998).

\bibitem{Landau}
L. Landau, ``On the theory of superfluidity of helium II'', J. Phys. U.S.S.R. {\bf 11}, 91 (1947).

\bibitem{Feynman}
R. P. Feynman, ``Atomic Theory of the Two-Fluid Model of Liquid Helium'', Phys. Rev. {\bf 94}, 262 (1954).

\bibitem{shlyap3}
A. Boudjem\^{a}a and G. V. Shlyapnikov, ``Two-dimensional dipolar Bose gas with the roton-maxon excitation spectrum'', Phys. Rev. A {\bf 87}, 025601 (2013).

\bibitem{snoke}
S. A. Moskalenko and D. W. Snoke, {\it Bose-Einstein Condensation of Excitons and Biexcitons}, Cambridge University Press (Cambridge), 2010.

\bibitem{microcavities}
A. V. Kavokin, J. J. Baumberg, G. Malpuech, and F. P. Laussy, {\it Microcavities}, Oxford University Press (Oxford), 2017.

\bibitem{lozovikIX}
Yu. E. Lozovik and V. I. Yudson, ``A new mechanism for superconductivity: pairing between spatially separated electrons and holes'', Sov. Phys. JETP {\bf 44}, 389 (1976).

\bibitem{jetpl0840222}
V. V. Solov'ev, I. V. Kukushkin, J. Smet, K. von Klitzing, and W. Dietsche, ``Kinetics of indirect electron-hole recombination in a wide single quantum
well in a strong electric field'', JETP Lett. {\bf 84}, 222 (2006).

\bibitem{shelykh1}
I. A. Shelykh, T. Taylor and A. V. Kavokin, ``Rotons in a Hybrid Bose-Fermi System'', Phys. Rev. Lett. {\bf 105}, 140402 (2010).

\bibitem{prb090165430}
A. K. Fedorov, I. L. Kurbakov, and Yu. E. Lozovik, ``Roton-maxon spectrum and
instability for weakly interacting dipolar excitons in a semiconductor
layer'', Phys. Rev. B {\bf 90}, 165430 (2014).

\bibitem{prl108060401}
M. Matuszewski, T. Taylor, and A. V. Kavokin, ``Exciton supersolidity in hybrid Bose-Fermi systems'', Phys. Rev. Lett. {\bf 108}, 060401 (2012).

\bibitem{prb091245302}
S. Yang, L. V. Butov, B. D. Simons, K. L. Campman, and A. C. Gossard, ``Fluctuation and commensurability effect of exciton density wave'', Phys.
Rev. B {\bf 91}, 245302 (2015).

\bibitem{prl098060405}
G. E. Astrakharchik, J. Boronat, I. L. Kurbakov, and Yu. E. Lozovik, ``Quantum phase transition in a two-dimensional system of dipoles'', Phys. Rev. Lett., ${\bf 98}$, 6, 060405 (2007).

\bibitem{pr0104000576}
O. Penrose and L. Onsager, ``Bose-Einstein condensation and liquid helium'', Phys. Rev. {\bf 104}, 576 (1956).

\bibitem{sci269000198}
M. H. Anderson, J. R. Ensher, M. R. Matthews, C. E. Wieman, and E. A. Cornell, ``Observation of Bose-Einstein condensation in a dilute atomic vapor'', Science {\bf 269}, 198 (1995).

\bibitem{jetpl0840329}
A. V. Gorbunov and V. B. Timofeev, ``Large-scale coherence of the Bose
condensate of spatially indirect excitons'', JETP Lett. {\bf 84}, 329
(2006).

\bibitem{butov2}
A. A. High, J. R. Leonard, A. T. Hammack, M. M. Fogler, L. V. Butov, A. V. Kavokin, K. L. Campman, and A. C. Gossard, ``Spontaneous coherence in a cold exciton gas'', Nature (London) {\bf 483}, 584 (2012).

\bibitem{dubin}
M. Alloing, M. Beian, M. Lewenstein, D. Fuster, Y. Gonz\'{a}lez, L. Gonz\'{a}lez, R. Combescot, M. Combescot, and F. Dubin, ``Evidence for a Bose-Einstein condensate of excitons'', Europhys. Lett. {\bf 107}, 10012 (2014).

\bibitem{rmp085000299}
I. Carusotto and C. Ciuti, ``Quantum fluids of light'', Rev. Mod. Phys. {\bf
85}, 299 (2013).

\bibitem{baumberg}
P. Cristofolini, G. Christmann, S. I. Tsintzos, G. Deligeorgis, G. Konstantinidis, Z. Hatzopoulos, P. G. Savvidis, J. J. Baumberg, ``Coupling Quantum Tunneling with Cavity Photons'', Science {\bf 336}, 704 (2012).

\bibitem{menon}
B. Datta, M. Khatoniar, P. Deshmukh, F. Thouin, R. Bushati, S. De Liberato, S. Kena Cohen and V. M. Menon, ``Highly nonlinear dipolar exciton-polaritons in bilayer MoS$_2$'', Nat. Commun. {\bf 13}, 6341 (2022)

\bibitem{tartakovskii}
C. Louca, A. Genco, S. Chiavazzo, T. P. Lyons, S. Randerson, C. Trovatello, P. Claronino, R. Jayaprakash, K. Watanabe, T. Taniguchi, S. Dal Conte, D. G. Lidzey, G. Cerullo, O. Kyriienko, and A. I. Tartakovskii, ``Nonlinear interactions of dipolar excitons and polaritons in MoS$_2$ bilayers'',  \url{arXiv:2204.00485} (2022).

\bibitem{cotlet}
O. Cotle\c{t}, S. Zeytino\v{g}lu, M. Sigrist, E. Demler, and A. Imamo\v{g}lu, ``Superconductivity and other collective phenomena in a hybrid Bose-Fermi mixture formed by a polariton condensate and an electron system in two dimensions'', Phys. Rev. B {\bf 93}, 054510 (2016).

\bibitem{sokolik}
A. Plyashechnik, A. A. Sokolik, N. S. Voronova, and Yu. E. Lozovik, ``Coupled system of electrons and exciton-polaritons: Screening, dynamical effects, and superconductivity'', \url{arXiv:2304.11245} (2023).

\bibitem{shelykh2}
O. Kyriienko and I. A. Shelykh, ``Elementary excitations in spinor polariton-electron systems'', Phys. Rev. B {\bf 84}, 125313 (2011).

\bibitem{prb062001548}
L. V. Butov, A. V. Mintsev, Yu. E. Lozovik, K. L. Campman, and A. C. Gossard,
``From spatially indirect excitons to momentum space indirect excitons by an
in-plane magnetic field'', Phys. Rev. B {\bf 62}, 1548 (2000).

\bibitem{prb062015323}
A. Parlangeli, P. C. M. Christianen, J. C. Maan, I. V. Tokatly, C. B.
Soerensen, and P. E. Lindelof, ``Optical observation of the energy-momentum
dispersion of spatially indirect excitons'', Phys. Rev. B {\bf 62}, 15323
(2000).

\bibitem{prl087216804}
L. V. Butov, C. W. Lai, D. S. Chemla, Yu. E. Lozovik, K. L. Campman, and A. C. Gossard, ``Observation of magnetically induced effective-mass enhancement of quasi-2D excitons'', Phys. Rev. Lett. {\bf 87}, 216804 (2001).

\bibitem{jetpl0890019}
A. V. Rossokhatyi and I. V. Kukushkin, ``Effect of the in-plane magnetic field on the recombination radiation spectrum of spatially-separated electron-hole layers'', JETP Lett., ${\bf 89}$, 19 (2009).

\bibitem{jetpl0890510}
A. V. Rossokhaty and I. V. Kukushkin, ``Dependence of the recombination
kinetics of spatially separated electron-hole layers of the parallel magnetic
field'', JETP Lett. {\bf 89}, 510 (2009).

\bibitem{gorkov}
L. P. Gor'kov and I. E. Dzyaloshinski\v{i}, ``Contribution to the theory of the Mott exciton in a strong magnetic field'', Sov. Phys. JETP {\bf 26}, 449 (1968).

\bibitem{ruvinskii_pla}
Yu. E. Lozovik, A. M. Ruvinsky, ``Magnetoexcitons in coupled quantum wells'', Phys. Lett. A {\bf 227}, 271--284 (1997).

\bibitem{prb065235304}
Yu. E. Lozovik, I. V. Ovchinnikov, S. Yu. Volkov, L. V. Butov, and D. S.
Chemla, ``Quasi-two-dimensional excitons in finite magnetic fields'', Phys.
Rev. B {\bf 65}, 235304 (2002).

\bibitem{tokatly_ssc}
A. A. Gorbatsevich and I. V. Tokatly, ``Formation of $k$-space indirect
magnetoexcitons in double-quantum-well direct-gap heterostructures'', Semicond. Sci. Technol. {\bf 13}, 288--295 (1998).

\bibitem{prb087205302}
G. J. Schinner, J. Repp, E. Schubert, A. K. Rai, D. Reuter, A. D. Wieck, A. O. Govorov, A. W. Holleitner, and J. P. Kotthaus, ``Many-body correlations of
electrostatically trapped dipolar excitons'', Phys. Rev. B {\bf 87}, 205302
(2013).

\bibitem{ssc144000399}
Yu. E. Lozovik, I. L. Kurbakov, G. E. Astrakharchik, J. Boronat, and M.
Willander, ``Strong correlation effects in 2D Bose-Einstein condensed dipolar
excitons'', Solid State Commun. {\bf 144}, 399 (2007).

\bibitem{prb080195313}
B. Laikhtman and R. Rapaport, ``Exciton correlations in coupled quantum wells
and their luminescence blue shift'', Phys. Rev. B {\bf 80}, 195313 (2009).

\bibitem{prb095245430}
Yu. E. Lozovik, I. L. Kurbakov, and P. A. Volkov, ``Anisotropic superfluidity of two-dimensional excitons in a periodic potential'', Phys. Rev. B {\bf 95}, 245430 (2017). 

\bibitem{semenov}
A. Semenov and Yu. Lozovik, ``On the superfluid properties of a polaritonic system'', Europhys. Lett. {\bf 78}, 67005 (2007).

\bibitem{jetpl0900146}
A. V. Gorbunov, V. B. Timofeev, D. A. Demin, and A. A. Dremin, ``Two-photon
correlations of luminescence at the Bose-Einstein condensation of dipolar
excitons'', JETP Lett. {\bf 90}, 146 (2009).

\bibitem{prb081235402}
T. Shi, L. Jiang, and J. Ye, ``Phase sensitive two-mode squeezing and photon
correlations from exciton superfluid in semiconductor electron-hole bilayer
systems'', Phys. Rev. B {\bf 81}, 235402 (2010).

\bibitem{prb075035303}
M. de Dios-Leyva, C. A. Duque, and L. E. Oliveira, ``Effects of crossed electric and magnetic fields on the electronic and excitonic states in bulk GaAs and GaAs/Ga$_{1-x}$Al$_x$As quantum wells'', Phys. Rev. B ${\bf 75}$, 035303 (2007). 

\bibitem{jetpl0830553}
V. V. Solov'ev, I. V. Kukushkin, J. Smet, K. von Klitzing, and W. Dietsche, ``Indirect excitons and double electron-hole layers in a wide single GaAs/AlGaAs quantum well in a strong electric field'', JETP Lett. {\bf 83}, 553 (2006).


\bibitem{prb046010193}
V. Srinivas, J. Hryniewicz, Y. J. Chen, and C. E. C. Wood, ``Intrinsic
linewidths and radiative lifetimes of free excitons in GaAs quantum wells'', Phys. Rev. B {\bf 46}, 10193 (1992).

\bibitem{deng2006}
H. Deng, D. Press, S. Gotzinger, G. S. Solomon, R. Hey, K. H. Ploog, and Y. Yamamoto, ``Quantum Degenerate Exciton-Polaritons in Thermal Equilibrium'', Phys. Rev. Lett. {\bf 97}, 146402 (2006).

\bibitem{prb104125301}
A. M. Grudinina, I. L. Kurbakov, Yu. E. Lozovik, and N. S. Voronova, ``Finite-temperature Hartree-Fock-Bogoliubov theory for exciton-polaritons'', Phys. Rev. B {\bf 104}, 125301 (2021).

\bibitem{e0NormalOrdering}
The quantity  $\epsilon_0[\hat{Q}^\dag(\boldsymbol{\rho})\hat{Q}(\boldsymbol{\rho})]$ in Eqs~(\ref{H}),~(\ref{U1B}) should be understood in the sense of the Taylor expansion of the function $\epsilon_0$ over its argument, where in every term of the series all $\hat{Q}^\dag(\boldsymbol{\rho})$ stand to the left-hand side with respect to all $\hat{Q}(\boldsymbol{\rho})$.

\bibitem{Bogoliubov}
N. Bogoliubov, On the theory of superfluidity, Acad. Sci. USSR. J. Phys. {\bf 11}, 23 (1947).

\bibitem{footnote2}
The kinetic function can be rewritten as $\mathcal{T}_{\bf p}\equiv p_x^2/2m_x+ p_y^2/2m_y + o({\bf p}^2)$, where the error $o({\bf p}^2)$ corresponds to interactions $U_0({\bf r})$ fully integrable at $r\to\infty$, while the axes orientation is chosen in a way such that the tensor of the quadratic part of $\mathcal{T}_{\bf p}$ is diagonal. For the condensate at rest (at ${\bf K}={\bf K}_0$), the masses are positively defined $m_x,m_y>0$ due to Eqs.~(\ref{Tp}), (\ref{T(p)}) and  the stability condition $\epsilon_0^{\prime\prime}(n_{\rm ex})>0$. Note that in the general case, due to $m_x\neq m_y$, polaritons possess anisotropic superfluidity (see Ref.~\cite{prb095245430}).

\bibitem{AGD}
A. A. Abrikosov, L. P. Gorkov, and I. E. Dzyaloshinskii {\it Methods of
Quantum Field Theory in Statistical Physics} (Dover, New York, 1975).

\bibitem{Popov}
V. N. Popov, {\it Functional integrals in quantum field theory and statistical physics}, Kluwer Academic Publishers (Dordrecht, Boston, Hingham, MA), 1983.

\bibitem{pr0155000080}
J. W. Kane and L. P. Kadanoff, ``Long range order in superfluid helium'',
Phys. Rev. {\bf 155}, 80 (1967).

\bibitem{prl121235702}
N. S. Voronova, I. L. Kurbakov, and Yu. E. Lozovik, ``Bose condensation of
long-living direct excitons in an off-resonant cavity'', Phys. Rev. Lett. {\bf 121}, 235702 (2018). 

\bibitem{prb103094511}
Yu. E. Lozovik, I. L. Kurbakov, G. E. Astrakharchik, and J. Boronat, ``Estimation of the condensate fraction from the static structure factor'',
Phys. Rev. B {\bf 103}, 094511 (2021).

\bibitem{tiop}
We take the Hamiltonian of photon leakout from the cavity in the form $\hat{\mathcal{H}}(t) = \sum_{\vec{q}}[L_{\vec{q}}^{\lambda}e^{-i\mu t/\hbar}\hat{c}_{\bf q}(t)\hat{c}_{\vec{q}}^{\dagger}(t) + \text{h.c.}]$ and assume instantaneous turn on of the decay at the moment of time $t=0$ and turn off at $t=t_0$. Then we calculate the dissipation rate at
$t_0\to\infty$ as $I_{\vec{q}}=(\hbar\omega_{\vec{q}}/t_0)\langle
\hat{S}^{\dagger}\hat{c}_{\vec{q}}^{\dagger}\hat{c}_{\vec{q}}\hat{S}
\rangle_0$, where $\hat{S}$ is the full $S$--matrix including $\hat{\mathcal{H}}(t)$ and all interactions, $\langle...\rangle_0$ is the bare average, and $\hat{c}_{\vec{q}}$ is the annihilation operator of a photon outside the cavity with the 3D momentum ${\vec{q}}$ and polarization corresponding to the condensate.

\bibitem{prl099140402}
M. Wouters and I. Carusotto, ``Excitations in a nonequilibrium Bose-Einstein
condensate of exciton polaritons'', Phys. Rev. Lett. {\bf 99}, 140402
(2007).

\bibitem{footnote1}
Coupling of light with the excitons on the second (non-condensed) bright branch may happen only within the polariton well. But thanks to extremely light polariton effective mass, the polariton well has a very small phase volume, hence the occupation of this branch is negligible~\cite{prb104125301}.

\bibitem{prb099085108}
N. A. Asriyan, I. L. Kurbakov, A. K. Fedorov, and Yu. E. Lozovik, ``Optical
probing in a bilayer dark-bright condensate system'', Phys. Rev. B {\bf
99}, 085108 (2019). 

\bibitem{combescot}
M. Combescot, O. Betbeder-Matibet, F. Dubin, ``The many-body physics of composite bosons'', Phys. Rep. {\bf 463}, 215--320 (2008).

\end{thebibliography}
\end{document}